\documentclass[useAMS, usenatbib]{mn2e}

\usepackage{graphicx}
\usepackage{amssymb,amsmath}
\usepackage{natbib}

\newcommand{\alfven}    {{Alfv$\acute{\rm e}$n }}
\newcommand{\beq}	{\begin{equation}}
\newcommand{\eeq}	{\end{equation}}
\newcommand{\beqa}{\begin{eqnarray}}
\newcommand{\eeqa}{\end{eqnarray}}
\newcommand{\alfvenic}  {{Alfv$\acute{\rm e}$nic}}
\newcommand{\beqs}	{\begin{displaymath}}
\newcommand{\eeqs}	{\end{displaymath}}
\newcommand{\beqas}	{\begin{eqnarray*}}
\newcommand{\eeqas}	{\end{eqnarray*}}

\def\bit{\begin{itemize}}
\def\eit{\end{itemize}}

\def\simlt{\lower.5ex\hbox{$\; \buildrel < \over \sim \;$}}
\def\simgt{\lower.5ex\hbox{$\; \buildrel > \over \sim \;$}}
\def\la{\simlt}
\def\ga{\simgt}

\font\tenbi=cmmib10 
\newfam\bifam \def\bi{\fam\bifam\tenbi} \textfont\bifam=\tenbi

\font\tenbr=cmbx10
\newfam\brfam  \textfont\brfam=\tenbr

\font\squinttenbi=cmbx10 at 9pt
\scriptfont\brfam=\squinttenbi

\def\vecnabla{
              \setbox1=\hbox{$\bigtriangledown$}
                           \raise.45ex\hbox{$\bigtriangledown$\hskip-.97\wd1
                           $\bigtriangledown$\hskip-.97\wd1
                           $\bigtriangledown$\hskip-.97\wd1}
                           \raise.47ex\hbox{$\bigtriangledown$}}

\def\vecv{{\bi{v}}}

\def\vecz{{\bi{z}}}

\def\vecB{{\bi{B}}}

\def\symbol#1{\ifmmode#1\else$#1$\fi}

\newcommand{\msun}{M_{\odot}}

\usepackage{hyperref}
\usepackage{deluxetable}

\title[Star Cluster Formation in Turbulent, Magnetized Dense Clumps]{Star Cluster Formation in Turbulent, Magnetized Dense Clumps with Radiative and Outflow Feedback}

\author[Myers et al.]
        {Andrew T. Myers$^1$,
        Richard I. Klein$^{2, 3}$, 
        Mark R. Krumholz$^4$, and
        Christopher F. McKee$^{1,2}$ \\
        $^1$  Department of Physics, University of California, Berkeley,
                   Berkeley, CA 94720; atmyers@berkeley.edu \\
         $^2$ Department of Astronomy, University of California, Berkeley,
                   Berkeley, CA 94720 \\
         $^3$ Lawrence Livermore National Laboratory, P.O. Box 808, L-23, Livermore, CA 94550 \\
         $^4$ Department of Astronomy and Astrophysics,
                   University of California, Santa Cruz, CA 95064}

\begin{document}

\label{firstpage}

\maketitle

\begin{centering}
\section*{Abstract}
\end{centering}

We present three \textsc{Orion} simulations of star cluster formation in a $1000$ $\msun$, turbulent molecular cloud clump, including the effects of radiative transfer, protostellar outflows, and magnetic fields. Our simulations all use self-consistent turbulent initial conditions and vary the mean mass-to-flux ratio relative to the critical value over $\mu_{\Phi} = 2$,  $\mu_{\Phi} = 10$, and $\mu_{\Phi} = \infty$ to gauge the influence of magnetic fields on star cluster formation. We find, in good agreement with previous studies, that magnetic fields corresponding to $\mu_{\Phi} = 2$ lower the star formation rate by a factor of $\approx 2.4$ and reduce the amount of fragmentation by a factor of $\approx 2$ relative to the zero-field case. We also find that the field increases the characteristic sink particle mass, again by a factor of $\approx 2.4$. The magnetic field also increases the degree of clustering in our simulations, such that the maximum stellar densities in the $\mu_{\Phi} = 2$ case are higher than the others by again a factor of $\approx 2$. This clustering tends to encourage the formation of multiple systems, which are more common in the rad-MHD runs than the rad-hydro run. The companion frequency in our simulations is consistent with observations of multiplicity in Class I sources, particularly for the $\mu_{\Phi} = 2$ case. Finally, we find evidence of primordial mass segregation in our simulations reminiscent of that observed in star clusters like the Orion Nebula Cluster. 

\section{Introduction}

\label{sec:intro}

Most stars form in groups \citep{lada2003, bressert2010}, but theoretical \citep[e.g.][]{shu1977, mckee2002, mckee2003} and numerical \citep{larson1969, banerjee2004, hennebelle2008, krumholz2007, krumholz2010, myers2011, cunningham2011, myers2013} treatments of star formation frequently consider stars forming in isolation. While these models are an important building block, they cannot capture the interaction effects likely to be important in real regions of star formation. For example, in \cite{krumholz2012}, who considered the collapse of a relatively massive ($1000$ $\msun$) molecular cloud clump, the presence of a few massive stars affected the temperature structure of the entire cluster. A true understanding of star formation requires considering the \emph{clustered} mode of formation commonly encountered in nature.  

Simulations of star cluster formation that include magnetic effects have typically ignored radiative transfer \citep{li2006, wang2010}, while simulations that include radiation have frequently ignored magnetic fields \citep{offner2009, krumholz2011, hansen2012, krumholz2012}. Important exceptions are \cite{price2009}, which studied the collapse of a $50$ $\msun$ molecular cloud including both magnetic and radiative effects, and \cite{peters2011}, which included magnetic fields and used a ray-tracing approximation for both the ionizing and non-ionizing components of the protostellar radiation. Non-zero field strengths can, among other things, reduce the overall rate of star formation \citep{price2009, padoan2011, federrath2012}, suppress fragmentation \citep{hennebelle2011, commercon2011, federrath2012, myers2013}, and influence the core mass spectrum \citep{padoan2007}, while radiative feedback is likely crucial to picking out a characteristic mass scale for fragmentation \citep{bate2009, myers2011, krumholz2011, krumholzorion2011, krumholz2012}. In this paper, we extend the work of \cite{krumholz2012} by including magnetic fields, and of \cite{price2009} by including self-consistently turbulent initial conditions, protostellar outflows, forming a statistically meaningful sample of stars, and following the protocluster evolution until a steady state is reached. The outline of this paper is as follows: we describe our numerical setup in Section \ref{sec:simulations}, report the results of our simulations in Section \ref{sec:results}, discuss our results in Section \ref{sec:discussion}, and conclude in Section \ref{sec:conclusions}.     

\section{Simulations}
\label{sec:simulations}

We have performed six simulations of star formation in turbulent molecular cloud clumps, aimed towards quantifying the effects of varying the magnetic field strength. The first three simulations have a maximum resolution of $\Delta x_f \approx 46$ AU and have either a strong, weak, or zero magnetic field. The next three are identical, except that the resolution is $\Delta x_f \approx 23$ AU instead. As the high-resolution simulations are necessarily more computationally expensive, we integrate them for a shorter period and use them mainly to check for convergence at early times. The parameters of all six runs are summarized in Table 1.

Our simulations consist of two distinct phases: a \emph{driving} phase, in which we generate turbulent initial conditions using a simplified set of physics, and a \emph{collapse} phase, in which we follow the gravitational collapse and subsequent star formation. In this section, we summarize our numerical approach and describe the initial conditions for each of these phases in turn. 

\subsection{Numerical Methods}

We use our code \textsc{Orion} to solve the equations of gravito-radiation-magnetohydrodynamics in the two-temperature, mixed-frame, gray flux-limited diffusion (FLD) approximation. \textsc{Orion} uses adaptive mesh refinement (AMR) \citep{berger1989} to focus the computational effort on regions undergoing gravitational collapse, and sink particles (\cite{lee2013}, see also \cite{bate1995}, \cite{krumholz2004}, \cite{federrath2010}) to represent matter that has collapsed to densities higher than we can resolve on the finest level of refinement. \textsc{Orion} uses Chombo as its core AMR engine, the HYPRE family of sparse linear solvers, and an extended version of the Constrained Transport scheme from PLUTO \citep{pluto} to solve the MHD sub-system \citep{li2012b}. The output of our code is the gas density $\rho$, velocity $\vecv$, magnetic field $\vecB$, the non-gravitational energy per unit mass $e$, the gravitational potential $\phi$, and the radiation energy density $E_R$, defined on every cell in the AMR hierarchy.  

The equations and algorithms that govern our simulations, as well as our choices of dust opacities, flux limiters, and refinement criteria, are with one exception identical to those in \cite{myers2013}. For a complete description of our numerical techniques, see that paper and the references therein. The exception is that, in the present work, we have also included the sub-grid protostellar outflow model of \cite{cunningham2011}. In short, in addition to accreting matter from the grid, the sink particles in these simulations also inject a portion of the accreted matter back to the simulation domain at high velocity in the direction given by the sink particle's angular momentum vector. Specifically, each sink ejects 21\% of the mass it accretes back into the gas at a velocity of 1/3 the Keplerian speed at the stellar surface, $v_{k, i} = \sqrt{G M_i / r_i}$, where $M_i$ and $r_i$ are the mass and radius of the $i$th sink particle. These parameters were selected so that the momentum flux would be consistent with observed values \citep{cunningham2011}, without the wind speed dominating the Courant time step. Additionally, the sub-grid outflow model employed in our calculations occasionally drives shocks strong enough to heat a small fraction of the gas to temperatures higher than the dust sublimation temperature ($\gtrsim$ $10^3$ K). Under such conditions, the dust opacity drops to nearly zero, and our normal treatment of the radiative transfer would not allow this gas to cool efficiently. Physically, this high-temperature gas should still cool by line emission and at still higher temperatures by radiation from free electrons, but it is difficult to model these processes using a single opacity. To remedy this, we make one further change from \cite{myers2013}: when the gas temperature $T_g$ in a cell exceeds $10^3$ K, we remove energy from that cell at a rate given by $(\rho/m_{\rm{H}})^2 \Lambda(T_g)$ and deposit it into the radiation field, where $m_{\rm{H}}$ is the hydrogen mass and $\Lambda(T_g)$ is the line cooling function from \cite{cunningham2006}. After the next radiation solve, this excess energy will be smoothed away by the diffusion solver. Without this correction, the temperature in these wind-shocked regions would be unphysically large.

We use periodic boundary conditions on all gas variables and on the gravitational potential $\phi$. The lone exception is the radiation energy density $E_R$. Periodic boundary conditions would trap radiation inside the simulation volume, which is not realistic. Instead, we use Marshak boundary conditions equivalent to surrounding the box in a radiation bath with temperature $T_r = 10$ K.

\subsection{Initial Conditions} 
\label{sec:ICs}

\begin{table*}
\begin{minipage}{175mm}
\caption{Simulation Parameters\label{runsetup}}
\centering
\begin{tabular}{@{}lcccccccccc}
\hline
Name & $\mu_{\Phi}$ & $B_0 $(mG) & $\beta_0$ & $\mathcal{M}_{A, 0}$ & $\mu_{\Phi, \rm{rms}}$ & $B_{\rm{rms}}$ (mG) & $\beta_{\rm{rms}} $ & $\mathcal{M}_{A, \rm{rms}}$ & $N_0$ & $\Delta x_f $ (AU) \\
\hline
Hydro & $\infty$ & 0.00 & $\infty$ & $\infty$ & $\infty$ & 0.00     & $\infty$ & $\infty$ & 128 & 46  \\
Weak & 10.0       & 0.16 & 0.24      & 3.8        &     2.8    & 0.57      & 0.02     &  1.1        & 128 & 46 \\
Strong  & 2.0        & 0.81 & 0.01     & 0.8       &     1.9     & 0.84      & 0.01     & 0.8         & 128 & 46  \\
Hydro$_{\rm{23}}$ & $\infty$ & 0.00 & $\infty$ & $\infty$ & $\infty$ & 0.00     & $\infty$ & $\infty$ & 256 & 23  \\
Weak$_{\rm{23}}$ & 10.0       & 0.16 & 0.24      & 3.8        &     2.8    & 0.57      & 0.02     &  1.1        & 256 & 23 \\
Strong$_{\rm{23}}$  & 2.0        & 0.81 & 0.01     & 0.8       &     1.9     & 0.84      & 0.01     & 0.8         & 256 & 23  \\
  \hline
 \end{tabular}
 
  \medskip
Col.\ 2: mass-to-flux ratio. Col.\ 3: mean magnetic field. Col.\ 4: mean plasma $\beta_0 = 8 \pi \rho c_s^2 / B_0^2$.  Col.\ 5: \alfven Mach number. Col. 6-9: same as 2-5, but for the root-mean-square field instead of $B_0$. Col.\ 10: resolution of the base grid. Col.\ 11: maximum resolution at the finest level. All runs have $M_c = 1000$ $\msun$, $L = 0.46$ pc, $\Sigma_c = 1$ g cm$^{-2}$, $\sigma_v = 1.2$ km s$^{-1}$, $\mathcal{M} = 11.1$, and 4 levels of refinement.
\end{minipage}
\end{table*}

We begin with a uniform, isothermal gas inside a box of size $L = 0.46$ pc. The initial gas temperature is $T_g = 10$ K and the initial density $\bar \rho$ is $6.96 \times 10^{-19}$ g cm$^{-3}$, or $n_{\rm{H}} = 2.97 \times 10^{4}$ hydrogen nuclei per cm$^{-3}$. The gravitational free-fall time
\beq
\label{tff}
t_{\rm{ff}} = \sqrt{\frac{3 \pi }{32 G \rho}}
\eeq computed using the mean density is $t_{\rm{ff}}(\bar \rho) \approx 80$ kyr. The corresponding total mass of the clump $M_c$ is $1000$ $M_{\odot}$, and the clump surface density $\Sigma_c = 1$  g cm$^{-2}$. 

These parameters were chosen to be consistent with observations of currently-forming star clusters that are large enough to contain high-mass stars \citep[e.g.][]{shirley2003, faundez2004, fontani2010} and are identical to those of \cite{krumholz2012}. In addition, our MHD runs have an initially uniform magnetic field with strength $B_0$ oriented in the $\vecz$ direction. The strength of this field can be expressed using the magnetic critical mass, $M_\Phi$, which is the maximum mass that can be supported against gravitational collapse by the magnetic field. In terms of the magnetic flux threading the box $\Phi = B_0 L^2$:
\beq
M_\Phi = c_{\Phi} \frac{\Phi}{G^{1/2}},
\eeq where $c_{\Phi} = 1 / 2 \pi$ for a sheet-like geometry \citep{nakano1978} and $\approx 0.12$ for a uniform spherical cloud \citep{mouschovias76, tomisaka88}. In this paper, we take $c_{\Phi} = 1 / 2 \pi$. The ratio of the mass in the box to the critical mass, $\mu_{\Phi} = M / M_{\Phi}$, thus divides the parameter space into magnetically sub-critical ($\mu_{\Phi} < 1$) cases, for which the field is strong enough to stave off collapse, and magnetically super-critical ($\mu_{\Phi} > 1$) cases, which will collapse on a timescale of the order of the mean-density gravitational free-fall time. Note that $\mu_{\Phi}$ here refers to the box as a whole, and not to the individual cores and clumps that form within.

Observations of the Zeeman effect in both OH lines \citep{troland2008} and CN $N = 1 \rightarrow 0$ hyperfine transitions \citep{falgarone2008} show that the typical value of $\mu_{\Phi}$ is $\approx 2$. While these observations do not rule out the existence of sub-critical magnetic fields in some star-forming regions, they do suggest that the typical mode of star formation involves fields that are not quite strong enough to support clouds by themselves over timescales longer than $\sim t_{\rm{ff}}(\bar \rho).$ Additionally, \cite{crutcher2010} suggest, based on a statistical analysis of observed line-of-sight magnetic field components, that values of $\mu_{\Phi}$ much more supercritical than $\mu_{\Phi} = 2$ may not be rare. In this paper we thus do two MHD runs, called Weak and Strong. Weak has an initial magnetic field strength of $B_0 = 0.16$ mG, corresponding to $\mu_{\Phi} = 10$. Strong, which is in fact closer to the mean observed $\mu_{\Phi}$, has $B_0 = 0.81$ mG and $\mu_{\Phi} = 2$. The corresponding values for the plasma parameter, $\beta_0 = 8 \pi \rho c_s^2 / B_0^2$, and the 3D \alfven Mach number, $\mathcal{M}_{A, 0} = \sqrt{12 \pi \rho} \sigma_v / B_0 $, where $\sigma_v$ is the 1D non-thermal velocity dispersion in the box, are shown in Table 1. We also do a run called Hydro, in which we set $B_0 = 0.0$ mG ($\mu_{\Phi} = \infty$).  Note that, because the Weak run is initially super-\alfvenic, there is some amplification of the initial magnetic field during the driving phase (see, e.g., \cite{federrath2011b}, \cite{federrath2011}). We thus also show in Table 1 the root-mean-squared magnetic field, $B_{\rm{rms}}$, as well as the values of $\mathcal{M}_A$, $\beta$, and $\mu_{\Phi}$ corresponding to $B_{\rm{rms}}$ instead of $B_{0}$. 

Molecular clouds and the clumps they contain are also observed to have significant non-thermal velocity dispersions \citep[e.g.][]{maclow2004, elmegreen2004, mckee2007, hennebelle2012}, which are generally explained by invoking the presence of supersonic turbulence. Turbulence is frequently modeled in simulations of star formation by generating a velocity field with the desired power spectrum (say, $P(k) \propto k^{-2}$ for supersonic Burgers turbulence) in Fourier space and then superimposing this field on a pre-determined smooth density distribution \cite[e.g.][]{krumholz2007b, bate2009, wang2010, girichidis2011, myers2013}. While this approach captures some of the effects of turbulence on cloud collapse, such as providing ``seeds" for fragmentation, it has the downside that density and velocity fields are not self-consistent at time $t = 0$. This lack of initial sub-structure in the density field permits collapse on the order of a free-fall time \citep{krumholz2012}. While this may be appropriate for simulations at the scale of individual cores, it is not appropriate for simulations at the scale of dense clumps or GMCs, as these structures convert only a small percentage of their mass to stars per free-fall time \citep{zuckerman1974, krumholztan2007, KDM2012, federrath2013}. Here, we instead follow the approach used in, e.g. \cite{klessen2000}, \cite{offner2009}, \cite{federrath2012}, \cite{hansen2012}, and \cite{krumholz2012}: we generate initial conditions using a driven turbulence simulation, and then switch on gravity and allow the gas to collapse. This ensures that the density and velocity fields are self-consistent at time zero. 

\begin{figure*}
\begin{minipage}{175mm}
  \centering
    \includegraphics[width=0.9\textwidth]{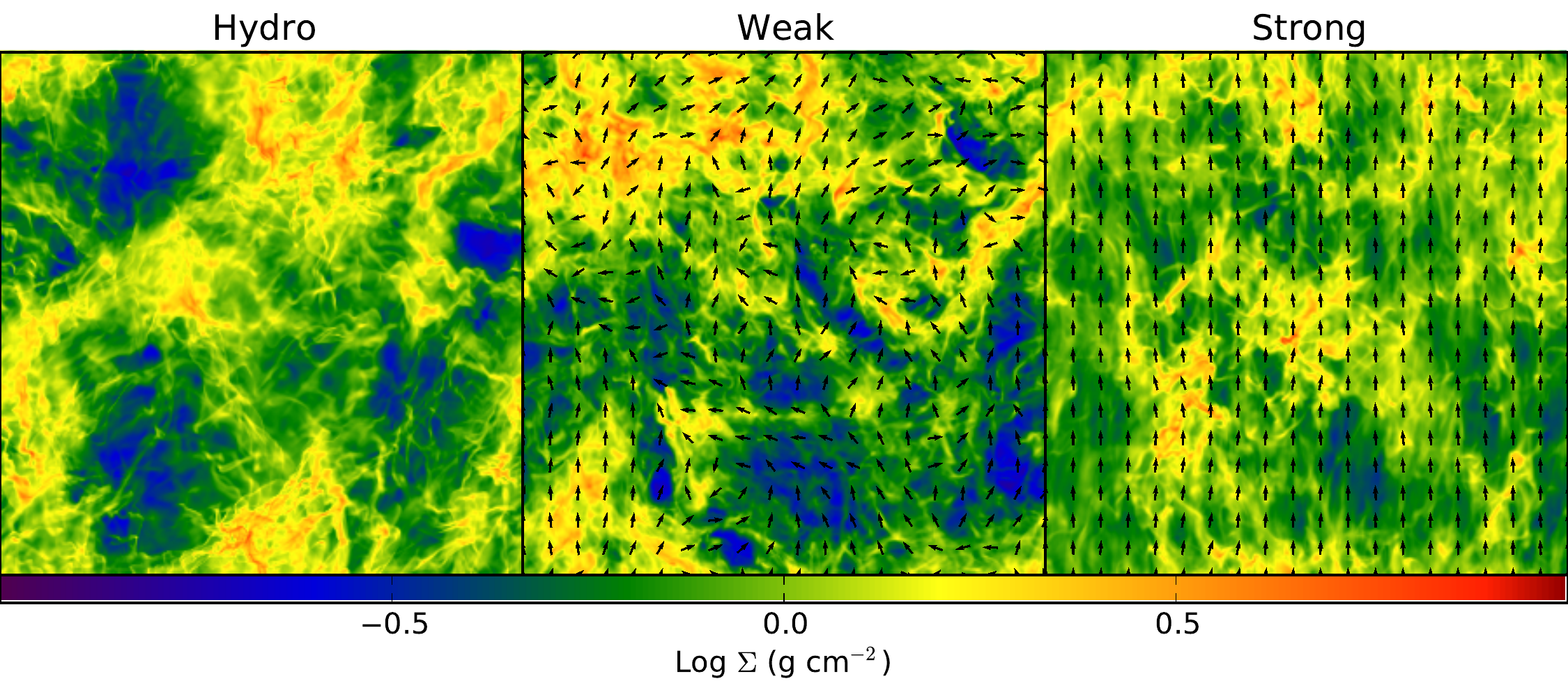}
    \caption{ \label{fig:initialconditions} Turbulent initial conditions for our three main runs. The colors indicate column density, while the mass-weighted, plane-of-sky magnetic field orientations are over-plotted as black arrows.}   
\end{minipage}
\end{figure*}

During the driving phase, we turn off self-gravity, particles, and radiative transfer, leaving just the ideal MHD equations. We set $\gamma = 1.0001$, so that the gas is very close to isothermal during this phase. For the driving pattern, we use a $512^3$ perturbation cube generated in Fourier space according to method in \cite{dubinski1995}. This pattern has a flat power spectrum in the range $1 \leq k L / 2 \pi \leq 2$, where $k$ is the wavenumber. We also perform a Helmholtz decomposition and keep only the divergence-free portion of the driving velocity, as in e.g. \cite{padoan1999}, \cite{ostriker1999, ostriker2001}, \cite{kowal2007}, \cite{lemaster2009}, and \cite{collins2012}. We then drive the turbulence using the method of \cite{maclow1999} for two crossing times. The resulting initial states for the collapse phase are illustrated in Figure \ref{fig:initialconditions}. Note that the initial conditions for Weak, for which $\mathcal{M}_A = 6.4$, contains much more structure in the magnetic field than those for Strong ($\mathcal{M}_A = 1.3$), in which the turbulence is not strong enough to drag around field lines significantly. 

Our choice of a solenoidal (divergence-free) driving pattern requires some discussion. The purpose of the driving is to mimic the effects of turbulence cascading down to our dense clump from larger scales. Since this is necessarily somewhat artificial, one would hope that the choice of driving pattern had little effect on the nature of the fully-developed turbulence. However, the presence of large-scale compressive motions in the driving has a significant effect on the density probability distribution function \citep{federrath2008}, the fractal density structure \citep{federrath2009}, and the star formation rate \citep{federrath2012}. The latter is of particular importance here. The turbulent runs in \cite{krumholz2012}, which used initial conditions quite similar to our Hydro run, had star formation rates that were too high by an order of magnitude. If the IMF peak is determined by the temperature structure imposed by protostellar accretion luminosities \citep{krumholz2011}, then overestimating the star formation rate likely means overestimating the characteristic stellar mass as well. Our choice of solenoidal driving helps bring the star formation rate closer to the observed values (Section 3.2), so the level of radiative feedback is probably more realistic in these calculations. Furthermore, even turbulence that is driven purely compressively will have approximately half the power in solenoidal modes in the inertial range for hydrodynamic, supersonic turbulence \citep{federrath2010b}, and magnetic fields further decrease the compressive fraction \citep{kritsuk2010, collins2012}. We thus expect that, whatever the driving mechanisms responsible for maintaining GMC turbulence on large scales, it would be mostly (but not purely) solenoidal by the time it cascades down to the $\approx 0.46$ pc scales of our box. At the end of the driving phase, our simulations have 29\%, 22\%, and 14\% of the total power in compressive motions in the Hydro, Weak, and Strong runs, respectively.     

After generating the initial conditions, we move on to modeling the collapse phase. We coarsen the turbulence simulations above from $N_0 = 512$ to either $N_0 = 256$ for the high-resolution runs or $N_0 = 128$ for our main runs. We switch on gravity, sink particles, and radiation, and also set $\gamma = 5/3$ instead of $\gamma = 1.0001$, appropriate for a gas of H$_2$ that is too cold for the rotational and vibrational degrees of freedom to be accessible. This also allows the temperature to vary according to the outcome of our radiative transfer calculation. We summarize the results of the collapse phase in the next section.

\section{Results}
\label{sec:results}

We begin by describing the evolution of the large-scale morphology of our clumps in section (Sec. \ref{sec:globalevolution}). We then discuss the overall rate of star formation (Sec. \ref{sec:star_formation}), compare our sink particle mass distributions to the stellar IMF (Sec. \ref{sec:IMF}) and to the protostellar mass functions of \cite{mckeeoffner2010} (Sec. \ref{sec:PMF}), examine the magnetic field geometry on the scale of individual stellar cores (Sec. \ref{sec:bfield}) and the accretion history of individual protostars (Sec. \ref{sec:TCAccretion}), describe the primordial mass segregation observed in our simulations (Sec. \ref{sec:mass_segregation}), and finally discuss the multiplicity of our simulated star systems (Sec. \ref{sec:multiplicity}). Unless otherwise stated, the results in this section are from our main Hydro, Weak, and Strong calculations at $\Delta x \approx 46$ AU. We discuss numerical convergence in section (Sec. \ref{sec:star_formation}).

\subsection{Global Evolution}
\label{sec:globalevolution}

\begin{figure*}
  \centering
    \includegraphics[width=0.9\textwidth]{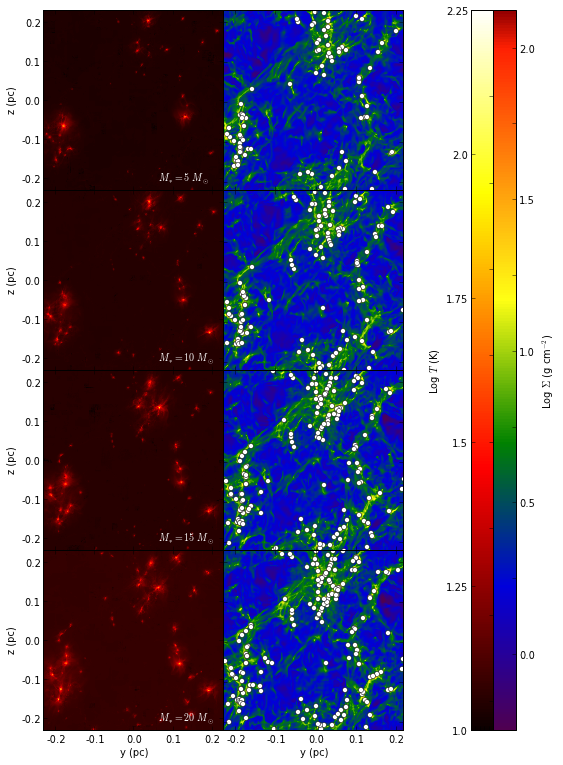}
    \caption{ \label{fig:HDpanels} Density-weighted mean temperature (left) and column density (right) for the Hydro run. Projected sink particle positions have been over-plotted as white dots.}   
\end{figure*}

\begin{figure*}
  \centering
    \includegraphics[width=0.9\textwidth]{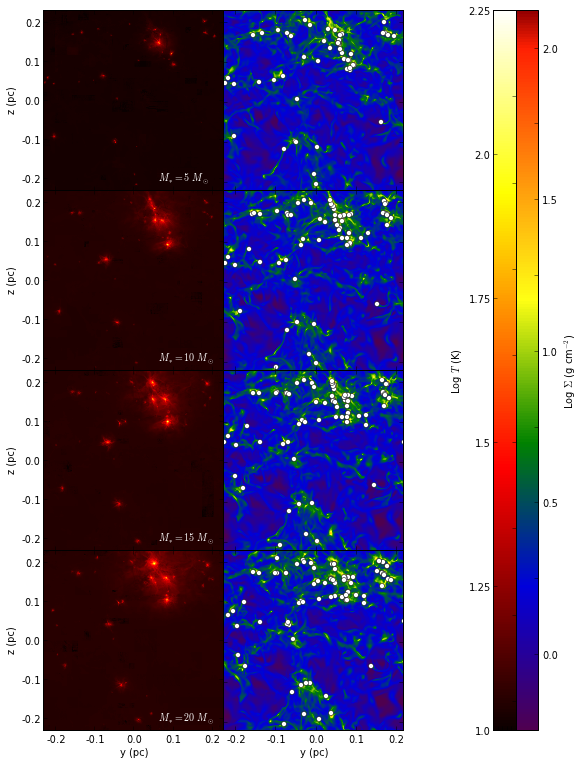}
    \caption{ \label{fig:Weakpanels} Same as Figure \ref{fig:HDpanels}, but for the Weak MHD run instead.}   
\end{figure*}

\begin{figure*}
  \centering
    \includegraphics[width=0.9\textwidth]{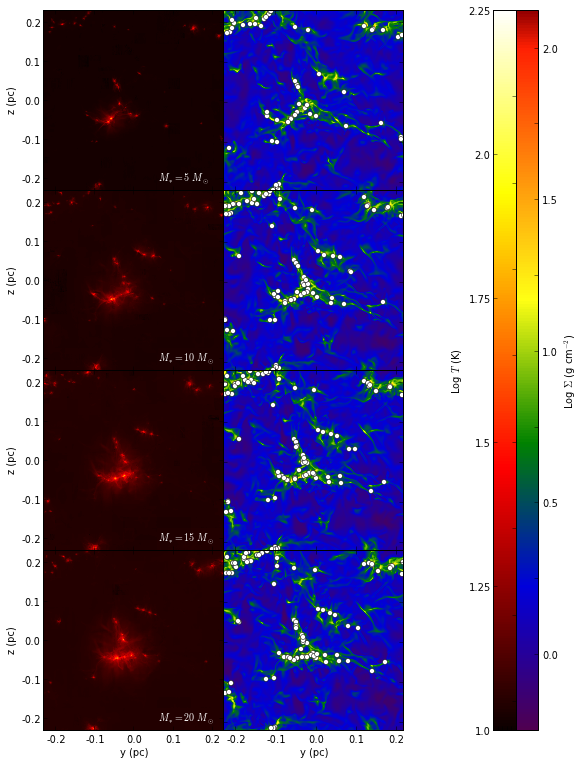}
    \caption{ \label{fig:Strongpanels} Same as Figure \ref{fig:HDpanels}, but for the Strong MHD run instead.}   
\end{figure*}

In Figures \ref{fig:HDpanels} through \ref{fig:Strongpanels}, we show the evolution of the column density $\Sigma$ and density-weighted mean gas temperature $T$, defined as $\Sigma = \int_{-L/2}^{L/2} \rho dx$ and $T = \int_{-L/2}^{L/2} \rho T_g dx / \Sigma$. Because star formation proceeds at different rates in the three runs (see Sec. \ref{sec:star_formation}) we compare the simulations based on the total mass that has been converted into stars, rather than the elapsed time. Figures \ref{fig:HDpanels} through \ref{fig:Strongpanels} show snapshots of the runs when the total mass in stars is 5, 10, 15, and 20 $M_{\odot} $. The global morphology of all three calculations is quite similar to the non-magnetic, turbulent simulations presented in \cite{krumholz2012}. In all three runs, the turbulence creates a network of over-dense, filamentary regions. As time passes, these dense regions collapse gravitationally and begin to fragment into isolated cores of gas. The cores collapse to form stars, leading to the appearance that stars tend to be strung along the gas filaments. Comparing the late-time distribution of stars in run Strong to those of run Weak and Hydro, two effects jump out. First, there are many more stars in Hydro than in the either Weak or Strong. Second, the magnetic field appears to confine star formation to take place within a smaller surface area in the $\mu_{\Phi} = 2$ case than in the others, so that the star particles tend to be found at higher surface density, and there are large regions with no stars at all. The reason for this behavior is simple: when the box as a whole is only magnetically supercritical by a factor of 2, then there are relatively large sub-regions within the domain that are magnetically sub-critical. These regions are not able to collapse to form stars on timescales comparable to $t_{\rm{ff}}$. We return to this point in section \ref{sec:mass_segregation}. 

\begin{figure*}
   \centering
     \includegraphics[width=0.9\textwidth]{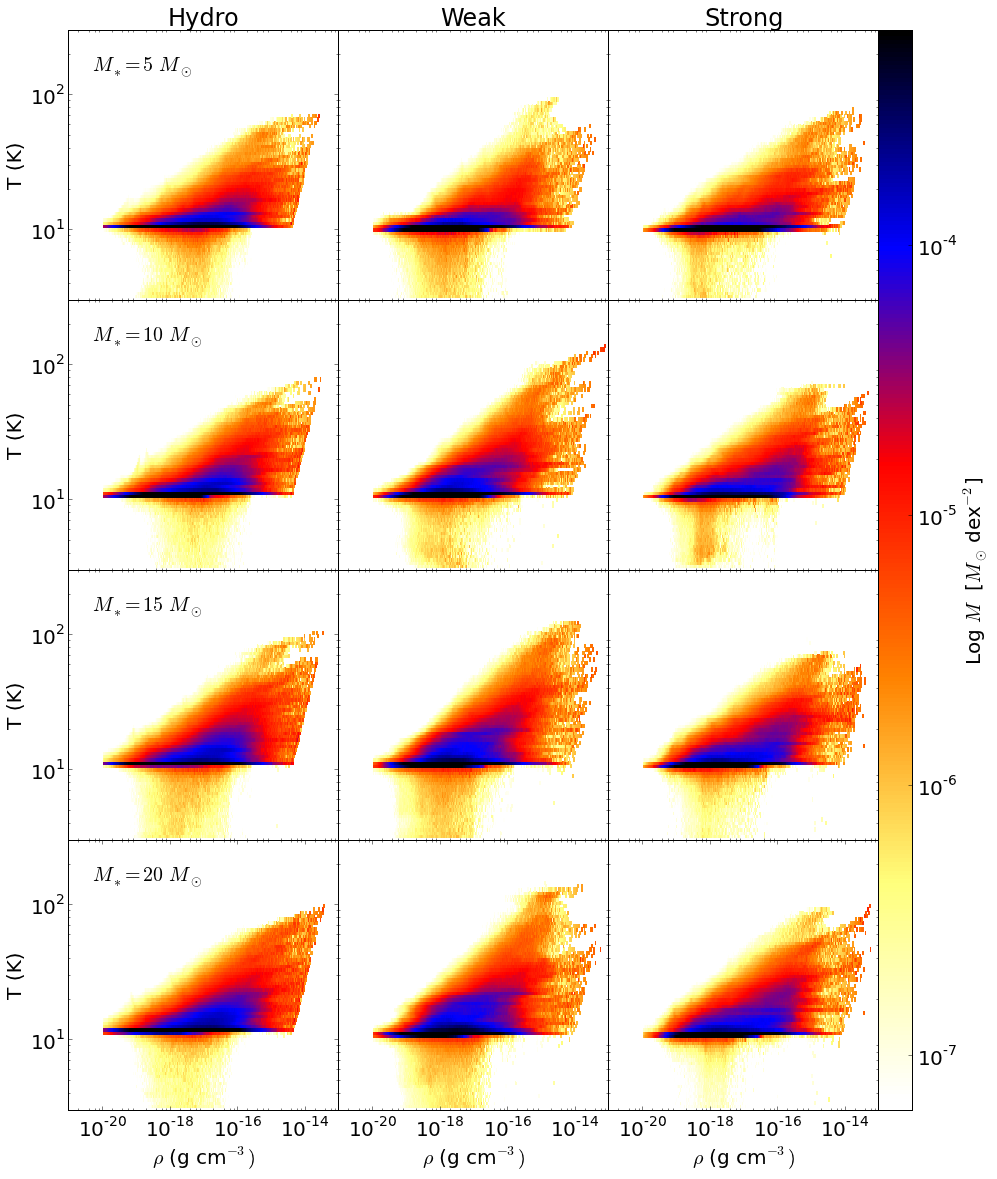}
     \caption{ \label{fig:PhaseMass} $T$-$\rho$ phase plots for all three runs. The columns, from left to right, correspond to the Hydro, Weak, and Strong runs, while the rows, from top to bottom, show the state of the simulations at the points at which 5, 10, 15, and 20 $M_{\odot}$ of stars have formed. The colors show the amount of mass in each $T$-$\rho$ bin.}   
\end{figure*}

\begin{figure*}
   \centering
     \includegraphics[width=0.9\textwidth]{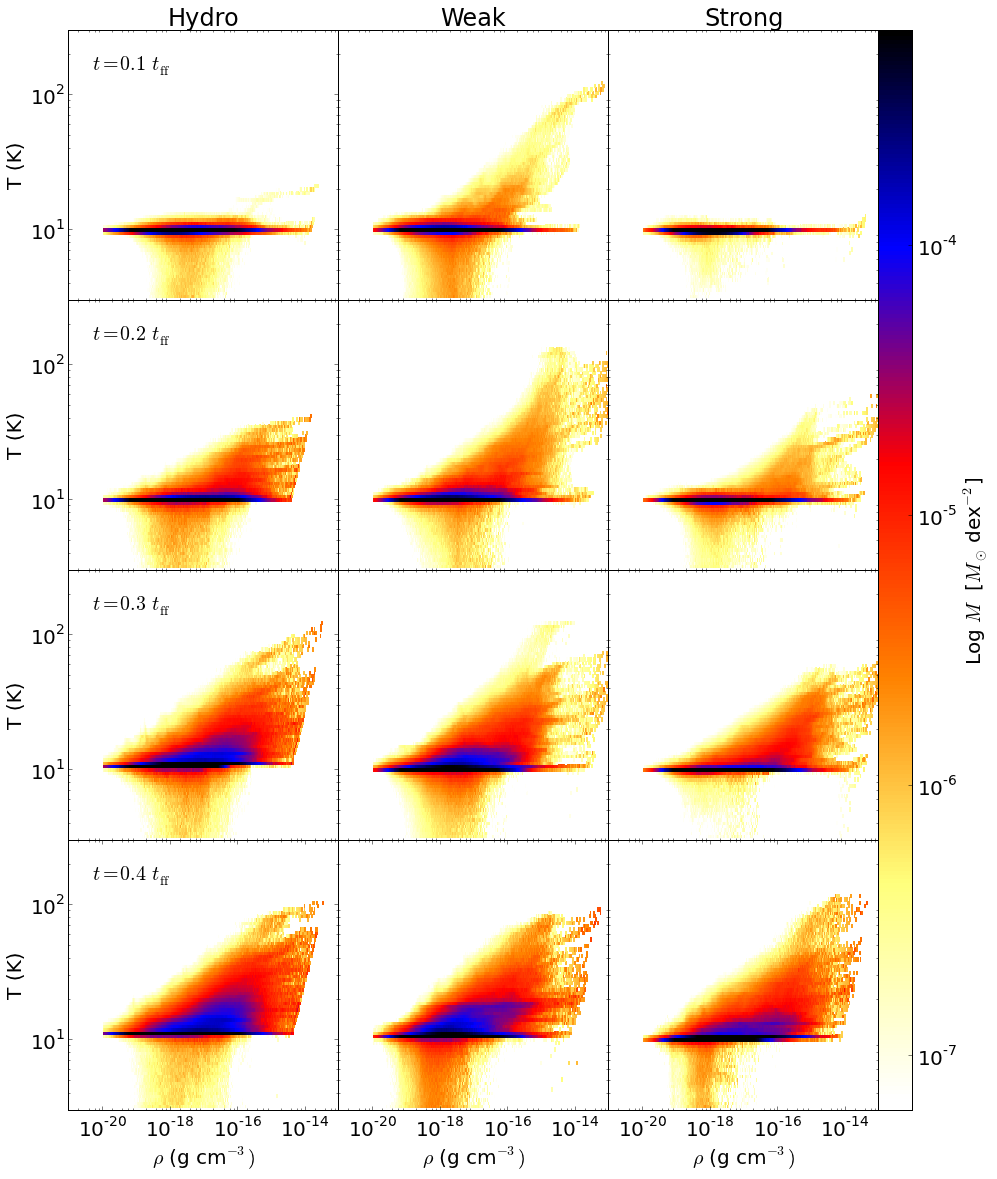}
     \caption{ \label{fig:PhaseTime} $T$-$\rho$ phase plots for all three runs. The columns, from left to right, correspond to the Hydro, Weak, and Strong runs, while the rows, from top to bottom, show the state of the simulations at $0.1$, $0.2$, $0.3$, and $0.4$ $t_{\rm{ff},\bar \rho}$ . The colors show the amount of mass in each $T$-$\rho$ bin.}   
\end{figure*}

The evolution of the gas temperature is interesting as well. At $t = 0$, the gas in the simulations is uniformly at 10 K. As stars form, they also heat up their surrounding environments. When the mass in stars is 5 $M_{\odot}$, the high-temperature regions are confined to the cores of gas around the individual protostars. As the simulations evolve and the protostars grow in mass, the heated regions grow and begin to overlap. By the time 20 $M_{\odot}$ of stars have formed, even regions far from any protostars have begun to be heated above the background temperature of 10 K, although the median gas temperatures are still a quite cool 11-12 K.          

We examine the temperature structure in our simulations more quantitatively in Figures \ref{fig:PhaseMass} and \ref{fig:PhaseTime}. These plots are constructed as follows. First, we create a set of 2-dimensional bins in $\rho - T_g$ space. We have chosen the bins to be logarithmically spaced in both $\rho$ and $T_g$, covering a range from $10^{-20}$ to $10^{-12}$ g cm$^{-3}$ in density and $10^{0.5}$ to $10^{2.5}$ K in temperature. Each bin is 0.025 dex wide in both $\rho$ and $T_g$, so that there are 320 density bins and 80 temperature bins. Then, we loop over every cell in the simulation. If a cell is not covered by a finer level of refinement (i.e. it is at maximum available resolution), we examine its density and temperature and add its mass to the appropriate bin. Otherwise, we skip it and move on. Figures \ref{fig:PhaseMass} and \ref{fig:PhaseTime} thus show the distribution of gas mass with both density and temperature, in units of $\msun$ dex$^{-2}$. 

We have performed this calculation for all three of our runs, comparing each at equal stellar masses (Figure \ref{fig:PhaseMass}) and at equal times (Figure \ref{fig:PhaseTime}). The differences between the three runs are particularly dramatic when the runs are compared at equal evolution times, because one of the effects of the magnetic field is to delay the rate of star formation (Section \ref{sec:star_formation}). However, even when compared at equal stellar mass (Figure \ref{fig:PhaseMass}), there is still less hot gas in the Strong run than the others. This is likely due to the overall lower accretion rate in the Strong run, since accretion luminosity is the dominant source of heating. The excess hot gas in the Weak run at early times is a small-sample size effect: there are only a few stars present at early times, and the Weak field run happens to form a few stars particularly early in its evolution (see Section \ref{sec:star_formation}). At later times, when there are dozens of stars in each run, the temperature structures of Weak and Hydro look quite similar.

\subsection{Star Formation}
\label{sec:star_formation}

\begin{figure}
  \centering
    \includegraphics[width=0.5\textwidth]{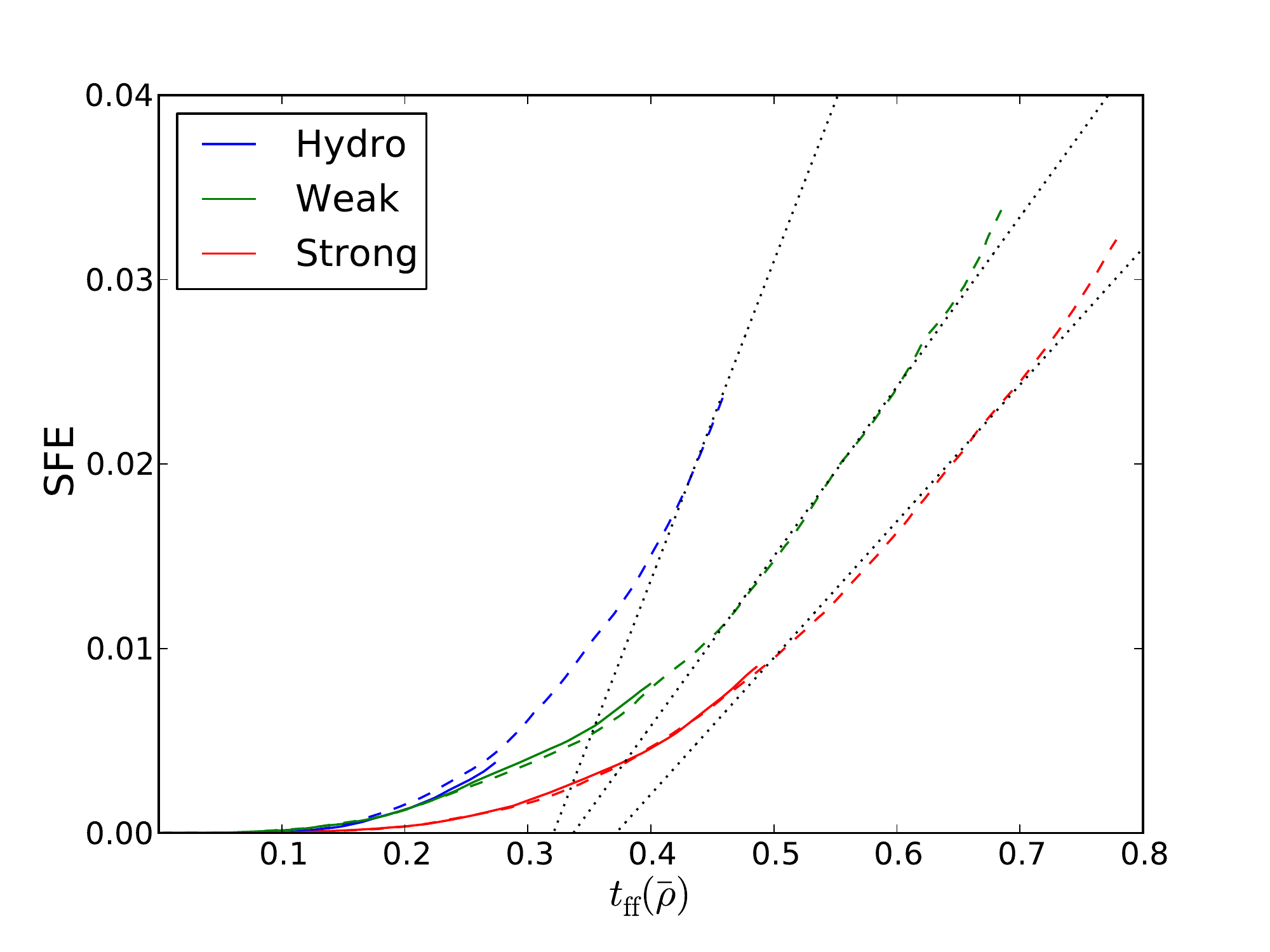}
    \caption{ \label{fig:StarM} Star formation efficiency (SFE) versus free-fall time for the Hydro (blue), Weak (green) and Strong (red) runs. The solid lines are from the high-resolution simulations, the dashed from the low. The black dotted lines demonstrate the slope of the low-resolution curves computed at SFE $= 0.02$, which we use to determine the star formation rate below. The free-fall time has been computed using the mean density.}   
\end{figure}

\begin{figure}
  \centering
    \includegraphics[width=0.5\textwidth]{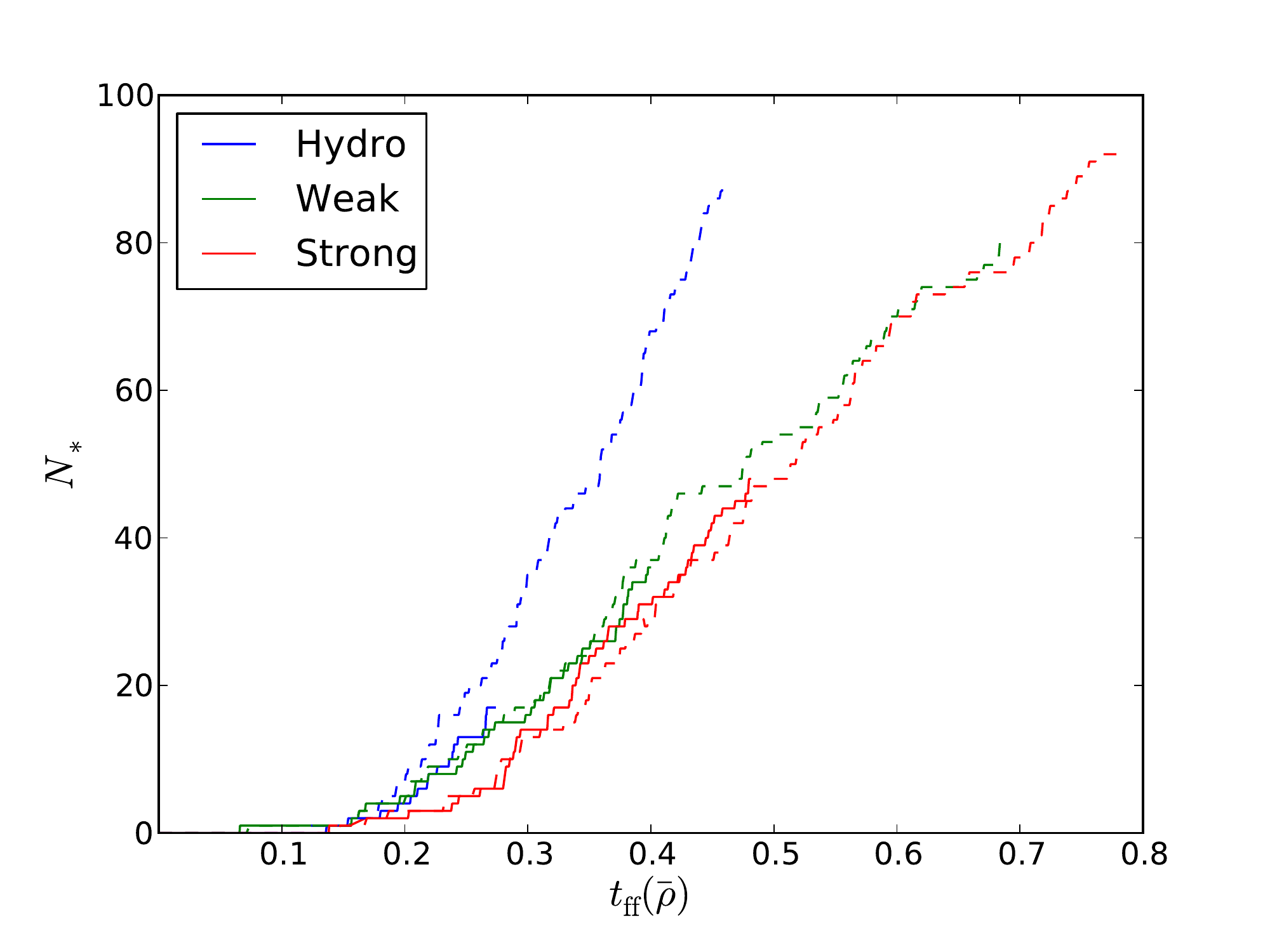}
    \caption{ \label{fig:StarN} Same as Figure \ref{fig:StarM}, but showing the number of stars, $N_*$, instead of the SFE.}   
\end{figure}

We now consider the properties of the sink particles formed in our simulations. In this section, we consider sink particles to be ``stars" when their masses exceed $0.05$ $\msun$. This threshold corresponds to the approximate mass at which second collapse occurs \citep{masunaga98, masunaga2000}. Below this mass, our code will merge sink particles if one enters the accretion zone of another, so only sinks with masses greater than $0.05$ $\msun$ are ensured to be permanent over the course of the simulations. With that caveat, we display the total number of stars $N_*$ and the star formation efficiency (SFE) versus time in our simulations in Figures \ref{fig:StarM} and \ref{fig:StarN}. We have taken the definition of the SFE to be the total mass in stars divided by the total mass of the cluster, including both gas and stars:
\beq
\rm{SFE} = \frac{{\it{M}_*}}{\it{M}_{\rm{gas}} + \it{M}_*} = \frac{\it{M}_*}{\it{M_c}} .
\eeq There is a monotonic decrease in both the SFE and $N_*$ at a given time with magnetic field strength. The reduction in $N_*$ between the $\mu_{\Phi} = \infty$ and $\mu_{\Phi} = 2$ cases is approximately a factor of 2. This agrees well with the simulations of \cite{hennebelle2011}, which found the same reduction in the number of fragments (a factor of $\approx$ 1.5 - 2 between $\mu_{\Phi} = 2$ and $\mu_{\Phi} = 120$) using quite different numerical schemes and initial conditions. For example, \cite{hennebelle2011} used an isothermal equation of state with a barotropic switch at high density, compared to our FLD radiative transfer, and took as initial conditions a spherical cloud with velocity perturbations, compared to our turbulent box initial conditions. This factor of $\approx 2$ also agrees with the isothermal calculations of \cite{federrath2012}, whose initial conditions were similar to our own. There is evidence from numerical simulations \citep{commercon2011, myers2013} that a combination of magnetic fields and radiative heating from accretion luminosity onto massive protostars can much more dramatically suppress fragmentation in the context  of massive ($\gtrsim 100$ $\msun$) core collapse, but as we do not form stars anywhere near as massive as those in \cite{myers2013} in these runs, this effect is not dramatic here. 

We also show in Figures \ref{fig:StarM} and \ref{fig:StarN} the SFE and $N_*$ versus free-fall time for the three high-resolution runs used in our convergence study. We find that as far as we have been able to run our high-resolution models, there is excellent convergence in the mass in stars as a function of time, and good convergence in $N_*$ as well. The largest discrepancy in $N_*$ occurs in the Hydro run at $t = 0.25$ $t_{\rm{ff}}$, when the low-resolution run contains $ \approx 50$\% more stars than the high-resolution run. An increase in the number of stars with decreasing resolution could be due to transient density fluctuations that exceed the threshold density for sink formation but do not truly lead to local collapse, as described in \cite{federrath2010}. This effect does not seem to be significant in either MHD run, likely because the density fluctuations in those cases are less extreme. 

Next, we examine the star formation rate (SFR) in our runs. The dimensional SFR, $\dot M_*$ is simply the rate at which gas is converted into stars, in e.g. $M_{\odot}$ yr$^{-1}$. There are various definitions of the \emph{dimensionless} SFR, $\epsilon_{\rm{ff}}$, in the literature; the most straightforward approach \cite[e.g.][]{krumholz2005, padoan2011, federrath2012} is to normalize $\dot M_*$ by what the star formation rate would be if all the mass in the box was converted to stars in a mean-density gravitational free-fall time:
\beq
\epsilon_{{\rm ff}, \overline{\rho}} = \frac{\dot{M}_*}{M_c / t_{\rm{ff}, \bar \rho}},
\eeq where $t_{\rm{ff}, \bar \rho}$ is Equation (\ref{tff}) evaluated at $\bar \rho$. However, because of the compressive effects of supersonic turbulence, most of the mass is actually at higher densities than $\bar \rho$. One could alternatively define some density threshold, $\rho_{\rm{thresh}}$, evaluate $t_{\rm{ff}}$ at that density, and define the relevant mass to be all the mass at $\rho_{\rm{thresh}}$ or higher. \cite{krumholz2012} take $\rho_{\rm{thresh}}$ to be the mass-weighted mean density, $\langle \rho \rangle$, and therefore define:
\beq
\epsilon_{{\rm ff}, \langle\rho\rangle} = \frac{\dot{M}_*}{(1/2) M_c / t_{\rm{ff}, {\langle \rho \rangle}}},
\eeq where $t_{\rm{ff}, {\langle \rho \rangle}}$ is the free-fall time (Equation \ref{tff}) computed using $\langle \rho \rangle$, and the factor of 1/2 accounts for the fact that, for a log-normal mass distribution, half the cloud mass is above $\langle \rho \rangle$. 

The first of these definitions is more analogous to extragalactic CO observations, in which the mass is taken to be all the mass in the beam, while the second is more analogous to observations of the SFR based on high-critical density tracers like HCN. We report both forms of $\epsilon_{{\rm ff}}$ in Table \ref{table:star_formation} below, where we have evaluated $\langle \rho \rangle$ at the instant gravity is switched on. Note that, while $\bar \rho$ is the same in all of our runs, $\langle \rho \rangle$ is not: the magnetic field keeps material from getting swept up across field lines, such that the value of $\langle \rho \rangle$ generally decreases with increasing magnetic field strength \citep{padoan2011}. 

The SFEs in Figure \ref{fig:StarM} are super-linear at all times. After about 0.2 $t_{\rm{ff}, \bar \rho}$, we find that the SFE versus $t$ curves are well-fit by power-laws of the form SFE$(t) \propto t^\alpha$, with $\alpha \approx 3.4,$, $2.7$, and $3.2$ for the Hydro, Weak, and Strong runs, respectively. This differs from the results of \cite{padoan2011} and \cite{federrath2013}, likely because unlike those authors we did not continue to drive the turbulence during the collapse phase. To compare the SFR across our runs, we compute and instantaneous $\dot M_*$ at the time at which the mass in stars is 20 $M_{\odot}$. This precise value is somewhat arbitrary, but it is consistent with observations of star-forming clouds, which generally have present-day SFEs of a few percent \citep{evans2009, federrath2013}. The resulting slopes are indicated by the dotted lines in Figure \ref{fig:StarM}. We summarize the values of $\dot M$, $\epsilon_{t_{\rm{ff}, \bar \rho}}$, and $\epsilon_{t_{\rm{ff}, \langle \rho \rangle}}$ in Table \ref{table:star_formation}. We find that the magnetic field decreases the SFR by a factor of $\approx 2.4$ over $\mu_{\Phi} = \infty$ to $\mu_{\Phi} = 2$, for both definitions of $\epsilon_{\rm{ff}}$. The $\approx 2.4$ reduction agrees well with previous studies of the SFR in turbulent, self-gravitating clouds \citep{price2009, padoan2011, federrath2012}. Likewise, our value of $\epsilon_{{\rm ff}, \overline{\rho}} = 0.17$ in the Hydro case is comparable to the value of 0.14 reported in the solenoidally driven, pure HD run in \cite{federrath2012}. This suggests that the radiative and outflow feedback processes included in this work have not dramatically altered the SFR over the time we have run, although a direct numerical experiment confirming this would be desirable. 

\begin{table}
\caption{ \label{table:star_formation} Summary of the star formation in each run.}
\centering
\begin{tabular}{@{}rcccccccccc}
\hline
Name & $t_{\rm{f}} / t_{\rm{ff}, \bar \rho}$ & $t_{\rm{ff}, \langle \rho \rangle}$ & $M_{*, t_f}$ & $N_{*, t_f}$ & $\dot M_*$ & $\epsilon_{{\rm ff}, \overline{\rho}}$ & $\epsilon_{{\rm ff}, \langle\rho\rangle}$ \\
\hline
Hydro  &$0.45$& 28.5 & 23.9 & 89 & 2.2 & 0.17 & 0.12\\
Weak &$0.68$ & 29.8 & 33.8 & 81 & 1.2 & 0.07 & 0.07\\
Strong &$0.78$ &33.4 & 32.2 & 92 & 0.9 & 0.07 & 0.05\\
  \hline
 \end{tabular}

\medskip
Col 2. - the final simulation time. Col. 3 - in kyr. Col 4. - the total mass in stars at $t_f$. Col 5. - the number of stars at $t_f$. Col. 6 - in $10^{-3}$ $M_{\odot}$ yr$^{-1}$.
\end{table}

Our Hydro run is almost identical to the ``TuW'' run from \cite{krumholz2012}. The exception is the turbulent driving pattern, which is solenoidal here and was a ``natural'' 1:2 mixture of compressive and solenoidal modes (i.e., 1/3 of the total power was in compressive motions) in \cite{krumholz2012}. \cite{federrath2012} found that mixed forcing increased the star formation rate by a factor of $\sim 3$ - 4 over the pure solenoidal case, depending on the random seed used to generate the driving pattern. If we compare our $\epsilon_{\rm{ff}, \langle \rho \rangle}$ to that of run TuW in \cite{krumholz2012}, we see that our driving pattern itself resulted in an $\approx 2.3$ reduction in the star formation rate, similar to the \cite{federrath2012} result. However, even with this reduction, the lowest SFR reported in this work of 0.05 in the Strong run is still slightly higher than the typically observed value of 0.01 \citep{krumholztan2007, KDM2012}. Likewise, \cite{federrath2013b} studied the dependence of the \cite{KDM2012} star formation law on the dimensionless SFR, finding that values of 0.003 to 0.04 covered range of scatter seen in the Milky Way and in other galaxies. Because of the sensitivity of the SFR to the details of the driving, which is in any event only a rough approximation to the true drivers of GMC turbulence, we believe that the raw numbers in Table \ref{table:star_formation} are to be taken less seriously than the trend with magnetic field strength, which appears to be robust for both solenoidal (this work, \cite{padoan2011}) and naturally mixed \citep{federrath2012} driving.   

\subsection{The Initial Mass Function}
\label{sec:IMF}

\begin{figure}
  \centering
    \includegraphics[width=0.5\textwidth]{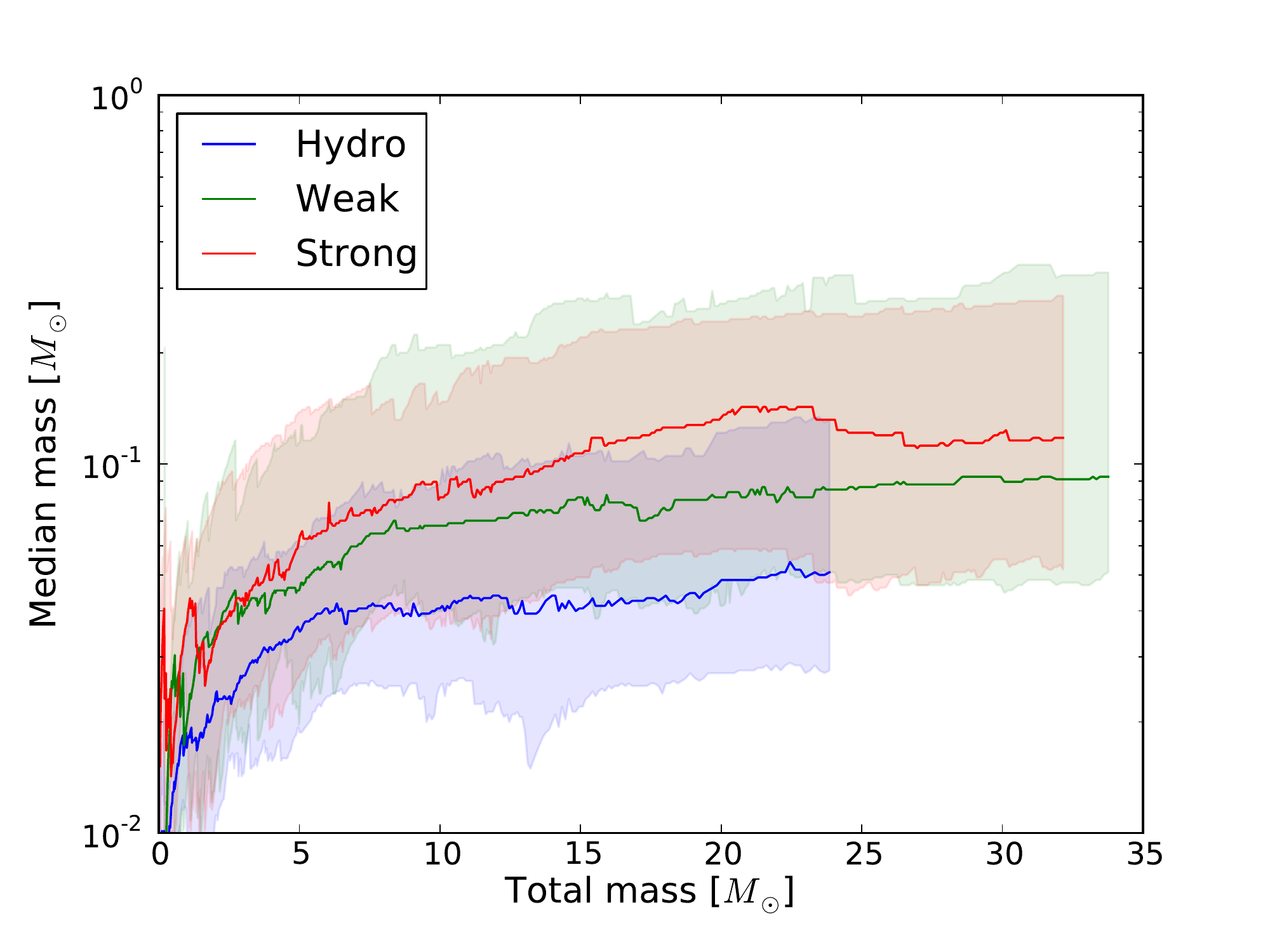}
    \caption{ \label{fig:median_mass} The solid lines refer to the median sink particle mass in the Hydro (blue), Weak (green), and Strong (red) runs. The shaded regions correspond to the middle 50\% of the sink particle mass distribution - i.e. the bottom edge of the shaded region traces out the 25th percentile mass and the top the 75th percentile mass.}   
\end{figure}

The stellar initial mass function (IMF) \citep[e.g.][]{chabrier2005} is one of the most basic observable properties of stellar populations, and serves as an important constraint on numerical simulations of star formation. In this section, we examine the distribution of sink particle masses in our simulations, and compare the result against the observed IMF. To begin, we show in Figure \ref{fig:median_mass} the evolution of the median, 25th percentile, and 75th percentile sink particle masses in each of our runs.

There are two points to make about this plot. First, in each of our runs, the 25th, 50th, and 75th percentile sink masses have all leveled off to well-defined values after about 10 to 20 $M_{\odot}$ of gas has been converted into stars. Even though most of the sink particles in our calculations are still accreting at the time we stop running, this growth is counterbalanced by the fact that new stars are continuously forming, so that the population as a whole has approached a steady-state distribution. Second, the median particle mass appears to monotonically increase with magnetic field strength, from $m_c = 0.05$ $M_{\odot}$ in the Hydro run to $m_c = 0.09$ $M_{\odot}$ in Weak and $m_c = 0.12$ $M_{\odot}$ in Strong. Thus, the magnetic field increases the median mass by a factor of $\approx 2.4$ over the range $\mu_{\Phi} = \infty$ to $\mu_{\Phi} = 2$. 

\begin{figure*}
  \centering
    \includegraphics[width=0.9\textwidth]{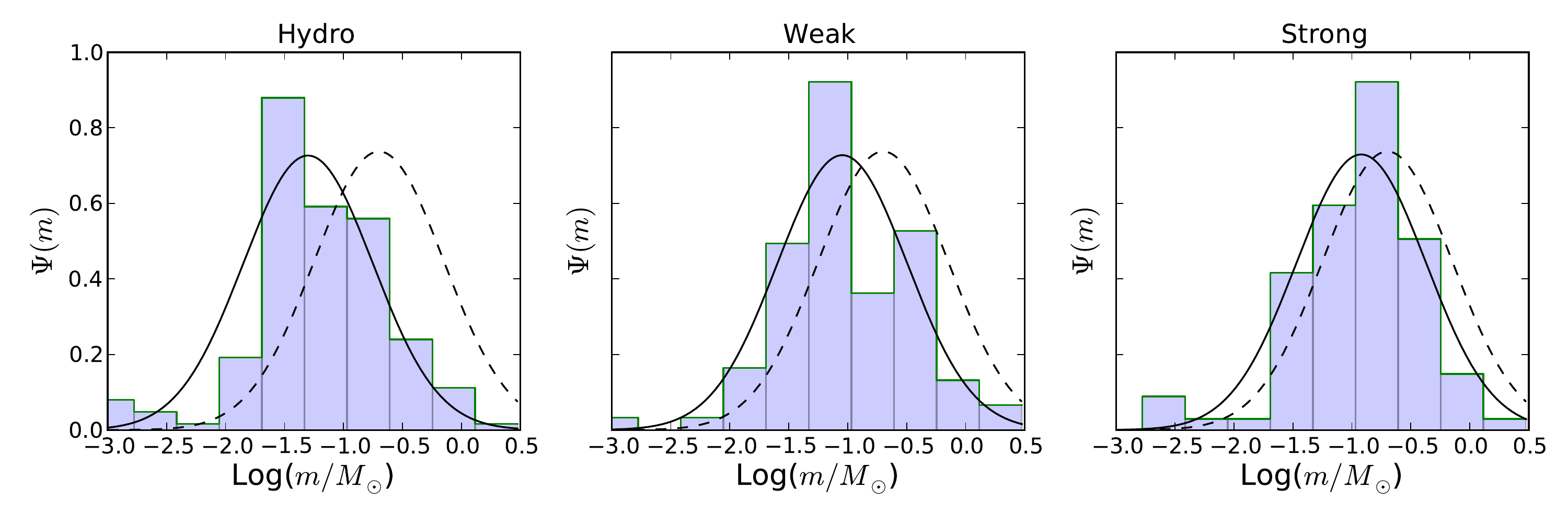}
    \caption{ \label{fig:IMF} Sink particle mass distributions for the Hydro, Weak, and Strong runs. The blue histograms are the simulation data, while the black solid and dotted lines are log-normal distributions (Equation (\ref{eq:lognormal})) with either the simulated value of $m_c$ (solid), or the Chabrier 2005 value (dotted).}
\end{figure*}

Next, we examine the full distribution of sink particle masses. In Figure \ref{fig:IMF}, we show the 
differential mass distribution, $\Psi(m)$, for each of our simulated clusters, where $\Psi(m)$ is defined such that $\int_{\log m_1}^{\log m_2} \Psi(m) d\log(m)$ gives the fraction of stars with $\log m$ between $\log m_1$ and $\log m_2$. We measure these functions at the point at which the total mass in stars is $20$ $M_{\odot}$. We find that the distributions are well-fit by a log-normal:
\beq
\label{eq:lognormal}
\Psi(m) \propto \exp[-\frac{(\log m - \log m_c)^2}{2 \sigma^2}],
\eeq where $m_c$ is the median mass for each run given above and $\sigma = 0.55$ is the log-normal width. If we take $m_c = 0.2$ $M_{\odot}$, this is equivalent to the low-mass limit of the \cite{chabrier2005} IMF.  However, even our Strong run, which has the largest $m_c$, is lower than the \cite{chabrier2005} characteristic mass by a factor of 1.7. The Weak and Hydro medians are smaller by factors of 2.2 and 4.0, respectively. 

\cite{mckee1989} considered an approximate expression for the maximum stable mass for a finite temperature cloud in the presence of magnetic fields:
\beq
\label{eq:mckee1989}
M_{\rm{cr}} \approx M_{\rm{BE}} + M_{\Phi},
\eeq where $M_{\rm{BE}} = 1.18 c_s^3 / \sqrt{G^3 \rho}$ is the Bonnor-Ebert mass and $M_{\Phi}$, as defined above, is the maximum stable mass for a pressureless cloud supported by magnetic fields. It is instructive to compute the typical value of $M_{\rm{cr}}$ in our simulations. If we write $M_{\rm{cr}} = \mu_{\Phi, \rm{core}} M_{\Phi}$, where $\mu_{\Phi, \rm{core}}$ is the mass-to-flux ratio at the \emph{core} scale, rather than the entire box, then:
\beq
M_{\rm{cr}} = \frac{\mu_{\Phi, \rm{core}}}{\mu_{\Phi, \rm{core}} - 1} M_{\rm{BE}}.
\eeq The Bonnor-Ebert mass for each of our runs, evaluated at the mass-weighted mean density, is 0.098 $\msun$, 0.102 $\msun$, and 0.114 $\msun$ for Hydro, Weak, and Strong, respectively. To estimate the value of $\mu_{\Phi, \rm{core}}$, we use $\mu_{\Phi, \rm{rms}}$. The resulting estimates for $M_{\rm{cr}}$ are 0.10 $\msun$, 0.16 $\msun$, and 0.23 $\msun$ - approximately a factor of 2 higher than our simulation results for the median stellar mass. The factor of $\sim 2.3$ increase from $\mu_{\Phi} = \infty$ to $\mu_{\Phi} = 2$ is quite close to the increase in $m_c$ we observe in our simulations. 

It is not surprising that our sink particles undershoot the \cite{chabrier2005} IMF somewhat - many of the sinks in Figure \ref{fig:IMF} formed only recently, and practically all of them are still accreting mass. The more relevant comparison is thus to the protostellar mass function (PMF), $\Psi_P$, in \cite{mckeeoffner2010}, which gives the mass distribution of a population of still-embedded Class 0 and I protostars. We compare our simulations to these theoretical PMFs in the next section. 

\subsection{The Protostellar Mass Function}
\label{sec:PMF}

The PMF depends on three factors: the functional form of $\dot{N}_*(t)$, the distribution of final stellar masses (i.e., the IMF), and the accretion history of the individual protostars, which can be calculated from various theoretical models of star formation. For example, competitive accretion (CA) \citep{Zinnecker1982, bonnell1997}, makes a different prediction about a star's accretion time as a function of its final mass than the turbulent core (TC) model of \cite{mckee2002, mckee2003}, so a population of still-accreting protostars with the same IMF and functional form of $\dot{N}_*(t)$ will have a different mass distribution under the two theories. 

\cite{mckeeoffner2010} provide PMFs for two functional forms of $\dot{N}_*$, one where it is constant and one where it exponentially accelerates with time.  In our simulations, we have a $N_*$ that is approximately linear with time (Figure \ref{fig:StarN}), at least after an initial transient phase of $\approx 0.2 t_{\rm{ff}}$, so we will not include any adjustments for accelerating star formation in our comparisons. We also do not include the ``tapered accretion" models considered in \cite{mckeeoffner2010}, as we find that the accretion rates in our simulations are well-described by non-tapered accretion (see Section \ref{sec:TCAccretion}). We have also followed \cite{mckeeoffner2010} in assuming that the distribution of final stellar masses $\Psi_C$ follows the \cite{chabrier2005} stellar IMF. We consider PMFs associated with three basic accretion models - the TC model, the CA model, and the isothermal sphere (IS) model of \cite{shu1977} - and two more complex models - two-component turbulent core (2CTC) model of \cite{mckee2003} and the two-component competitive accretion (2CCA) model of \cite{mckeeoffner2011}. 2CTC is a generalization of the TC model that limits to IS for low masses and TC for high masses, while 2CCA similarly interpolates for the IS and CA models. Having fixed $\Psi_C$ and the form of $\dot{N}_*(t)$, the only other parameter that enters into the ``basic" PMFs is the upper mass limit of stars that will form in the cluster, $m_u$. In our comparison, we set $m_u = 6$ $\msun$, which is larger than the most massive protostar we form in these simulations and about the mass of the most massive core identified in section \ref{sec:bfield}. The 2CTC model contains an additional parameter: the ratio $R_{\dot m}$ of the accretion rate for the TC model to the that of the IS model, evaluated for a 1 $\msun$ star. The 2CCA model contains a similar ratio between the CA and IS accretion rates at 1 $\msun$. We have taken $R_{\dot m} = 3.6$ for 2CTC and $R_{\dot m} = 3.2$ for 2CCA, which correspond to the fiducial parameter choices in \cite{mckeeoffner2010} and \cite{mckeeoffner2011}.

\begin{figure*}
  \centering
    \includegraphics[width=0.9\textwidth]{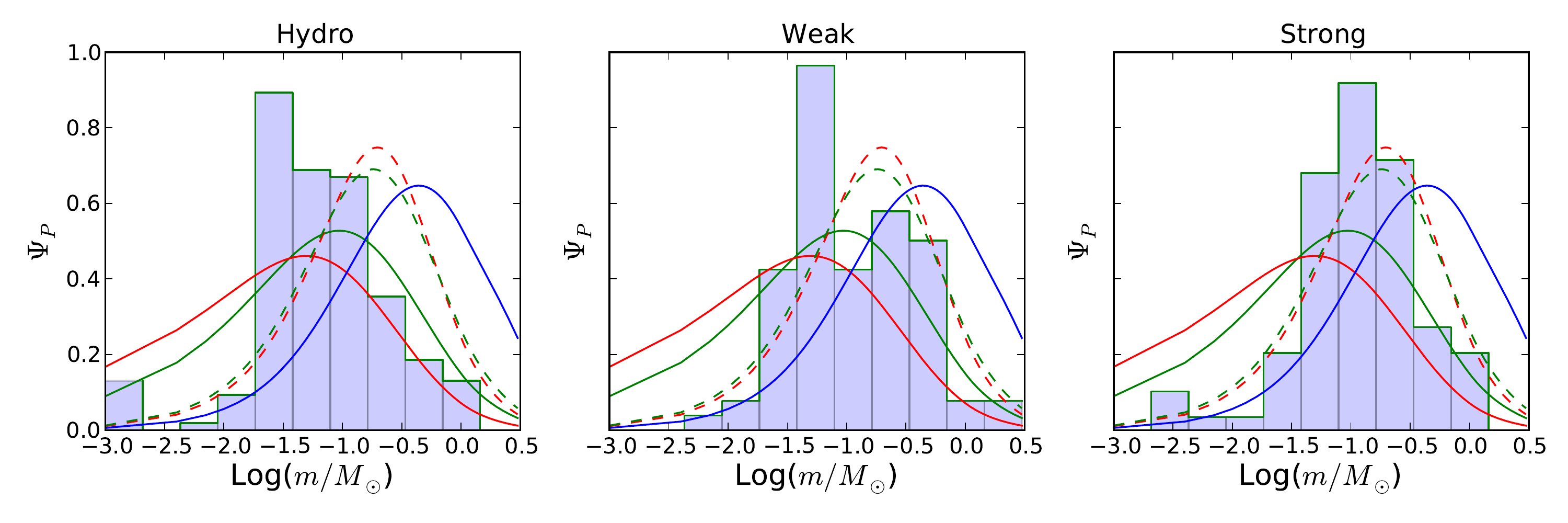}
    \caption{ \label{fig:PMFs} Protostellar mass distributions in our simulations at $M_* \approx 20 $  $\msun$ compared to the theoretical PMFs in McKee \& Offner 2011. The blue histograms are the simulation data. The green solid curve is the PMF associated with the TC model, the red solid curve the CA model, and the blue solid curve is the IS model. The green and red dotted curves are the 2CTC and 2CCA PMFs, respectively.}   
\end{figure*}

We show the distribution of protostellar masses for the Hydro, Weak, and Strong models in Figure \ref{fig:PMFs}. To make a clean comparison across the three runs, we have plotted the results at the times for which the total mass in stars is 20 $M_{\odot}$, or SFE $= 0.02$. The earliest time this occurs is $t \approx 0.4$ $t_{\rm{ff}}$ in the Hydro run, so we are well-outside the initial ``turn-on" phase during which the assumption of constant  $\dot{N}_*$ is inappropriate. We also show in Figure \ref{fig:PMFs} the five theoretical PMFs from above. The TC and CA models seem to predict more low-mass protostars than we find in our simulations, and the IS model predicts a median mass that is too large by about 0.5 dex. The two-component models, however, agree well with the median mass found in our Strong simulation. The Hydro run does not compare well with any of the theoretical models, mainly because its median mass is too low - lower than the Weak run by a factor of $\approx 2$ and the Strong run by a factor of $\approx 3$ for this snapshot. This increase in the typical protostellar mass due to the magnetic field appears to be necessary to get good agreement with the two-component PMFs.

\begin{figure}
  \centering
    \includegraphics[width=0.5\textwidth]{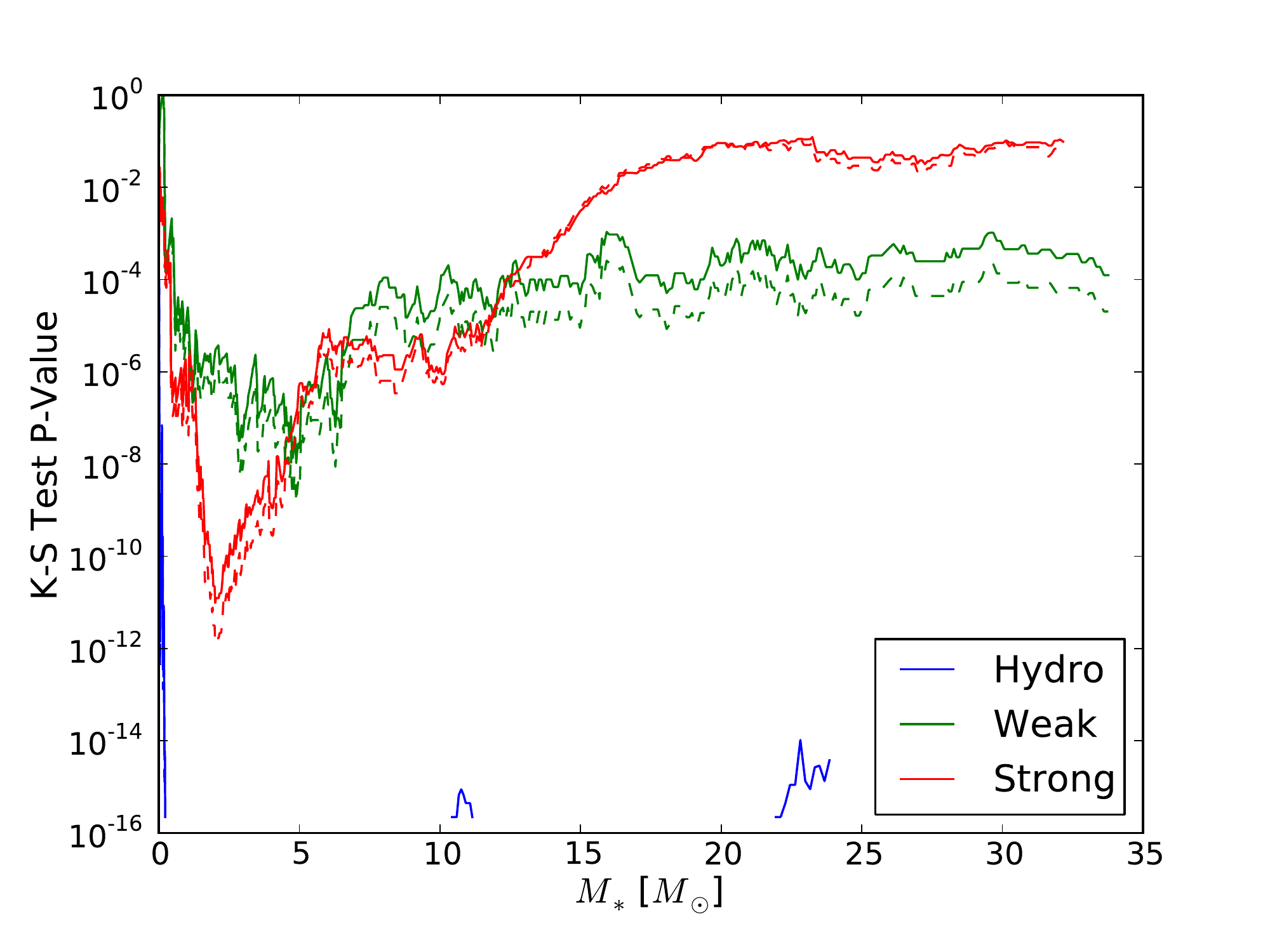}
    \caption{\label{fig:ks} K-S test results comparing our simulated protostar populations to the 2CTC (solid) and 2CCA (dotted models. The y-axis shows the $p$-value returned by the test, and the x-axis shows time.}   
\end{figure}

To examine the degree of agreement with the two-component PMFs over the entire evolution of the cluster, we perform a Kolmogorov-Smirnov (K-S) test comparing our simulated protostar populations to the 2CTC and 2CCA PMFs for all our data outputs. The results are shown in Figure \ref{fig:ks}. The Strong MHD run, after the initial transient phase, attains statistical consistency with both PMFs. This agreement appears to be steady with time, hovering around a K-S $p$-value of 0.1. The $p$-value for the Weak run is also relatively stable, although the agreement with the predicted PMFs is not as good. The Hydro distribution never reaches a steady $p$-value $> 10^{-4}$ for any of the models we consider. Note that, as the PMFs predicted by the 2CTC and 2CCA models are quite similar, our simulated PMFs cannot be said to favor one accretion history model over the other.    

\subsection{Core Magnetic Field Structure}
\label{sec:bfield}

\begin{figure*}
  \centering
    \includegraphics[width=0.9\textwidth]{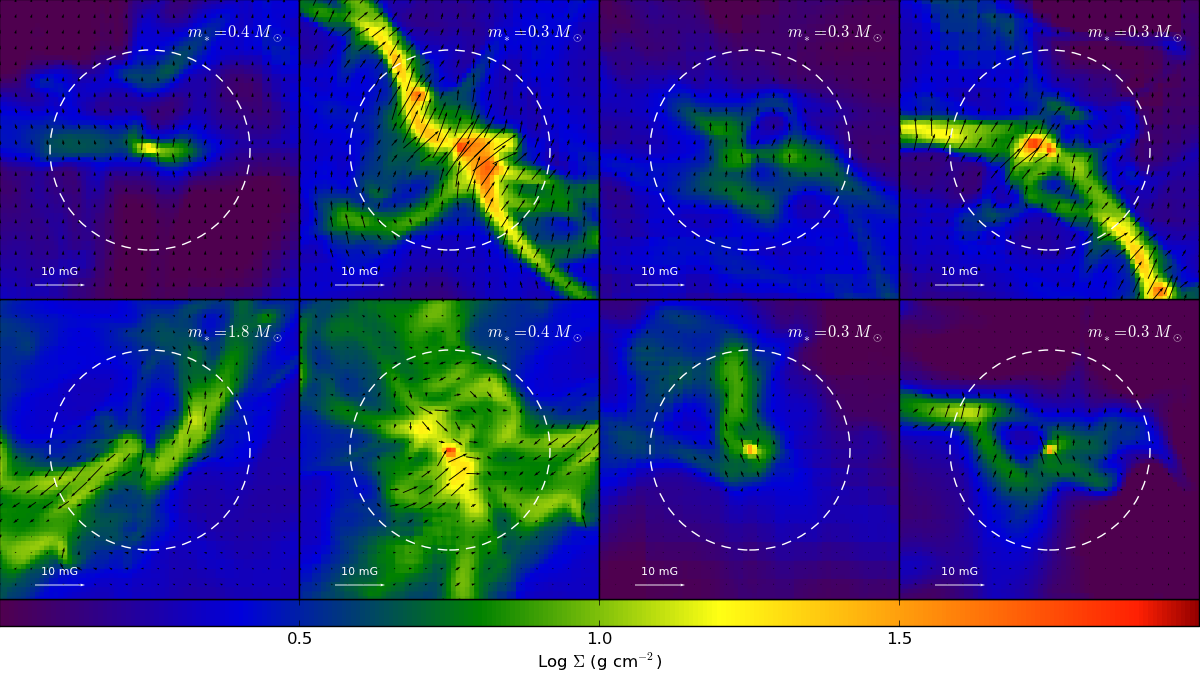}
    \caption{  \label{fig:cores} Top - zoomed in views of the four most massive protostars in the Strong field calculation at $t = 0.4$ $t_{\rm{ff}}$. The window size has been set to 3000 AU. The color scale shows the logarithm of the column density, and the black arrows show the mass-weighted, plane-of-sky magnetic field vectors. The masses of the protostars have been indicated in each panel. Bottom - same, but for the Weak MHD run.}   
\end{figure*}

\begin{figure*}
  \centering
    \includegraphics[width=0.9\textwidth]{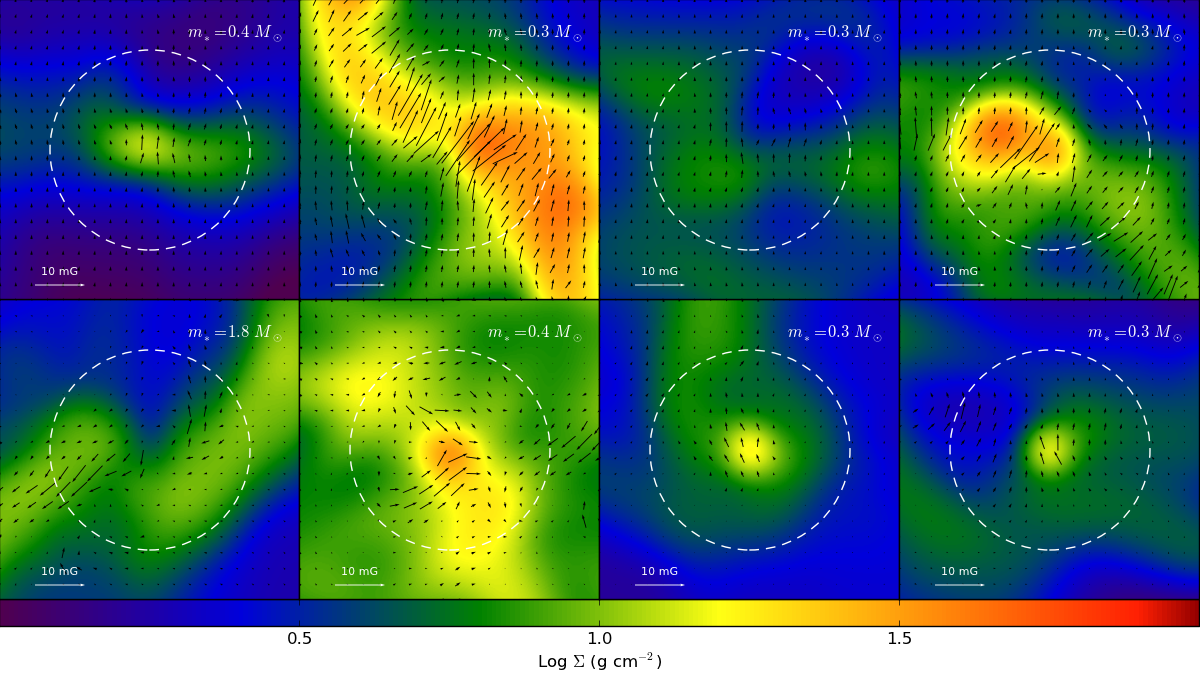}
    \caption{  \label{fig:cores1000} Same as Figure \ref{fig:cores}, except convolved with a 1000 AU Gaussian beam. The size of the beam is indicated by the white circle.}   
\end{figure*}

It is also useful to examine the geometry of the magnetic field in the cores formed in our simulations. From the Weak and Strong field MHD runs, we select the four most massive protostars at the time $t = 0.4 t_{\rm{ff}}$. These range in mass from 0.3 to 1.8 $\msun$ at this point in the calculation. In Figure \ref{fig:cores}, we show column density maps overlaid with density-weighted, projected magnetic field vectors showing the central 3000 AU surrounding each protostar. Figure \ref{fig:cores1000} shows the same cores convolved with a 1000 AU Gaussian beam to ease comparison with observations. As in \cite{krumholz2012}, we find that all the protostars are found near the centers of dense structures similar to the cores identified in dust thermal emission maps. The typical size of the cores, by inspection, is about 0.005 pc. We calculate the core mass by adding up all the mass (in gas and in the central sink particle) within a sphere of 0.005 pc radius around the protostar. The resulting core masses range from about 2 $\msun$ to 6 $\msun$. 

In the Strong run, we find that the magnetic field geometry always follows the ``hourglass" structure commonly observed in regions of low-mass \citep{girart2006, rao2009} and high-mass \citep{girart2009, tang2009} star formation. We see examples of this in the Weak case (the left two panels of Figure \ref{fig:cores}) but we also see examples of highly disordered field geometry (the right two panels). This is in part due to the greater ability of the protostellar outflows to disrupt magnetic field lines in the Weak field case. Note that, because our wind model caps the wind velocity at 1/3 the Keplerian value, this tendency for the winds to disrupt the field lines is if anything underestimated in our simulations. 

In general, dust polarization maps of star-forming cores tend to reflect magnetic fields that are quite well-ordered. If \cite{crutcher2010} is correct, and cores with $\mu_{\Phi} \gtrsim 10$ are not rare, then chaotic magnetic field geometries like those shown in the bottom panels of \ref{fig:cores} should not be rare, either. \cite{crutcher2010} argues for a flat distribution of field strengths from approximately twice the median value down to very near 0 $\mu$G. If this is true, and the median field corresponds to $\mu_{\Phi} = 2$, then a flat distribution implies that $\approx 10$\% of cores should have $\mu_{\Phi} \ge 10$.

\subsection{Turbulent Core Accretion}
\label{sec:TCAccretion}

\begin{figure*}
  \centering
    \includegraphics[width=0.9\textwidth]{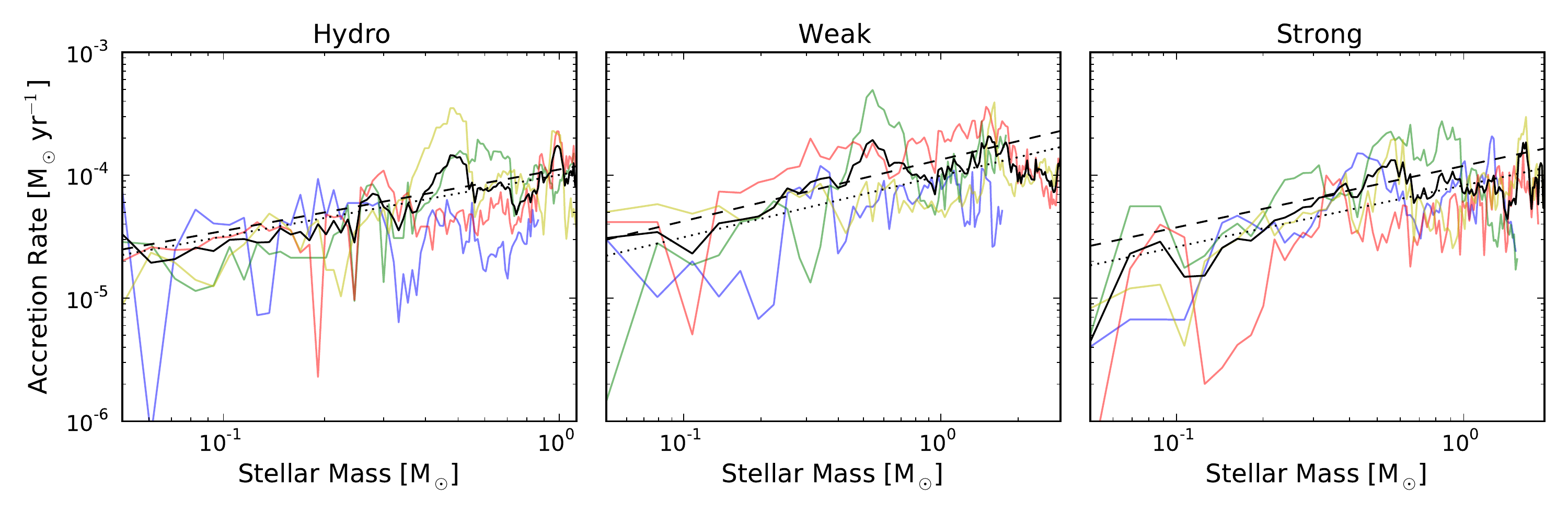}
    \caption{\label{fig:accretion} Plots of $\dot{m}_*$ versus $m_*$ for the four most massive protostars in each run. The solid colored lines correspond to the individual protostars, while the black solid line is the average $\dot{m}_*$ of all four. The black dashed line is the theoretical prediction of the TC model (see text), while the black dotted line is the best $\dot{m}_* \propto m_*^{1/2}$ power-law fit to the data. These lines overlap almost exactly for the Hydro case. The accretion rates have been smoothed over a 500-year timescale for clarity. }   
\end{figure*}

The turbulent core (TC) model of \citeauthor{mckee2002} (\citeyear{mckee2002, mckee2003}) is a generalization of the singular isothermal sphere \citep{shu1977} that was developed in the context of massive stars. In this model, both the gravitationally bound clump of gas where a cluster of stars is forming and the cores that form individual stars and star systems are assumed to be supersonically turbulent. The predicted accretion rate in the TC model is:
\beq
\label{TCaccretion}
\dot{m}_* = 1.2 \times 10^{-3} \left( \frac{m_{*,f}}{30 \hspace{3 pt} M_{\odot} } \right)^{3/4} \Sigma_{\rm{cl}}^{3/4} \left( \frac{m_*}{m_{*,f}} \right)^{1/2} \hspace{3 pt} M_{\odot} \hspace{3 pt}  \rm{yr}^{-1},
\eeq where we have increased the normalization constant by a factor of 2.6 from the \cite{mckee2003} value to account for subsonic contraction, as per \cite{tan2004}. In the above equation, $\dot{m}_*$ is the instantaneous mass accretion rate onto the protostar, $m_*$ is the protostar's current mass, $m_{*,f}$ is the final mass of the star once it is done accreting, and $\Sigma_{\rm{cl}}$ is the surface density of the surrounding molecular clump, which we identify with the mean surface density in our simulations, $\Sigma_c = 1$ g cm$^{-2}$. 

The TC model includes the effects of magnetic fields in an approximate way. Its prediction is that the effect of the field strength on the accretion rate should be quite modest. The value quoted above takes the magnetic field into account for a ``typical" field strength, for which $\mathcal{M}_A$ is $\approx 1$. According to \cite{mckee2003}, the accretion rate in the field-free case would be only $\approx$ 6\% higher, assuming that $\alpha_{\rm{vir}}$ is kept constant as the magnetic field strength is varied. 

To test this, we select the four most massive stars (as these are the stars for which the TC model should be most applicable) at the end of our Hydro, Weak, and Strong runs, and plot $\dot{m}_*$ versus $m_*$ over the accretion history of the protostars. We compare the simulation results to Equation \ref{TCaccretion}. As we also hold $\alpha_{\rm{vir}}$ constant across our runs, the TC model predicts that Equation \ref{TCaccretion} with the stated normalization should be quite accurate for all the runs, whatever the field strength. To estimate $m_{*,f}$, we take the sink particle mass at the most evolved time and add in all the gas remaining in the surrounding 1000 AU core, although this neglects material entrained by outflows and potential competition with nearby partners. The resulting average $m_{*,f}$ over the four most massive protostars is 2.0 $\msun$ for Hydro, 4.2 $\msun$ for Weak, and $2.6 \msun$ for Strong. 

The result is shown in Figure \ref{fig:accretion}. Overall, our simulation results agree quite well with the TC model, both in terms of the predicted power-law slope, $\dot{m}_* \propto m_*^{1/2}$, and the predicted normalization. There is a noticeable reduction in the accretion rate relative to Equation \ref{TCaccretion} with magnetic field strength, but overall this effect is small compared to the size of the fluctuations in the simulation data. To characterize the error in the TC prediction, we fit a $\log \dot{m}_* = C + (1/2) \log m_*$ power-law to the mean accretion rates for the four-star sample in each run (the solid black curves in Figure \ref{fig:accretion}). The resulting fits are compared against Equation (\ref{TCaccretion}) in Figure \ref{fig:accretion}. We find that the normalization of the best-fit power-law, $C$, is lower than the prediction of Equation (\ref{TCaccretion}) by 12\% in the Hydro run, 35\% in the Weak MHD run, and 44\% in the Strong MHD run. It is not surprising that the measured accretion rates in the magnetic cases differ somewhat from the prediction in Equation (\ref{TCaccretion}), since the latter is based on the assumptions that (1) the \alfven Mach number is unity in the star-forming cores, whereas we set only the initial $\mathcal{M}_A$ in the entire turbulent box; and (2) the mass-to-flux ratio in the star-forming cores is similar to that estimated by \cite{li1997}. 

 \subsection{Mass Segregation}
 \label{sec:mass_segregation}
 
 \begin{figure}
  \centering
    \includegraphics[width=0.5\textwidth]{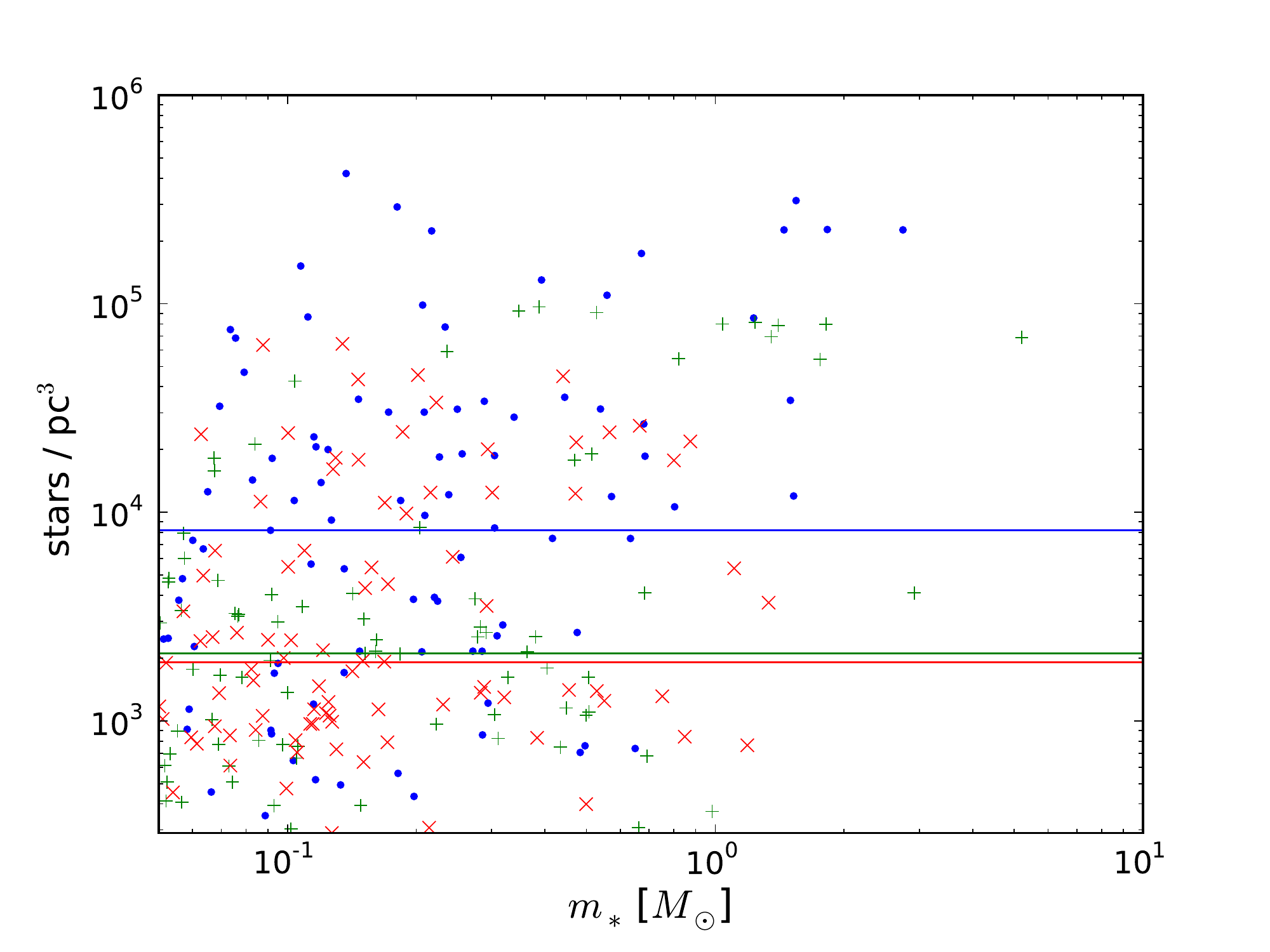}
    \caption{\label{fig:ms_volume} Stellar density $n_*$ versus stellar mass $m_*$ for all the stars in our simulations. The blue circles correspond to the Hydro run, the green plus symbols to Weak, and the red crosses to Strong. The colored lines show the median stellar density in each run.}   
\end{figure}

 \begin{figure}
  \centering
    \includegraphics[width=0.5\textwidth]{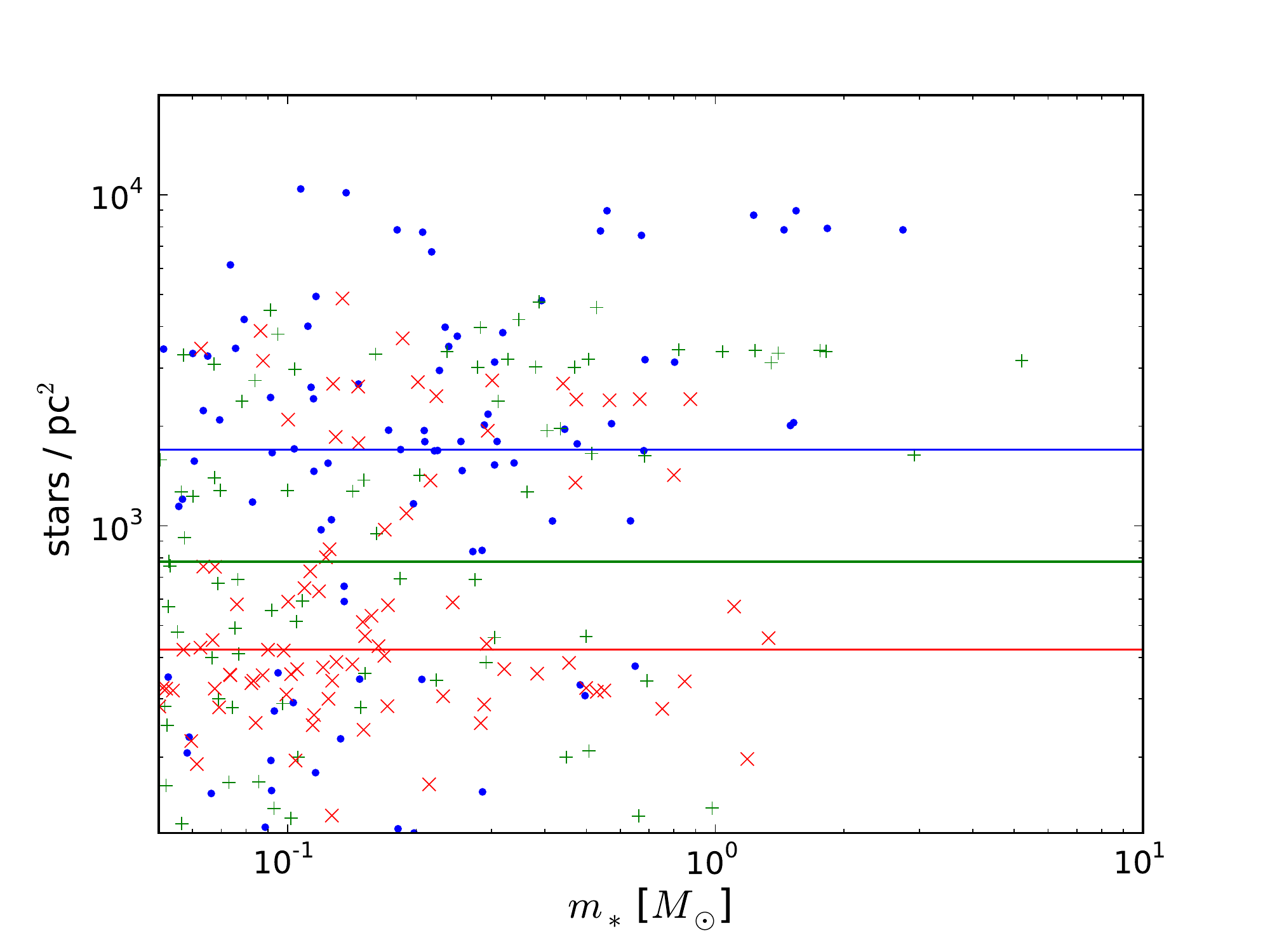}
    \caption{\label{fig:ms_surface} Same as Figure \ref{fig:ms_volume}, but for the stellar surface density $N_*$ instead.}   
\end{figure}
 
Much of our knowledge of the detailed inner structure of star clusters comes from studies of the Orion Nebula Cluster (ONC), which at $\sim 400$ pc is close enough to Earth to be easily observable. One interesting property of the ONC is that, with the exception of relatively massive stars like those that comprise the Trapezium, stars appear to be distributed throughout the cluster independently of mass.  However, stars more massive than $\approx 3$ $\msun$ appear to only be found in the center of the cluster, where the stellar surface density is highest \citep{huff2006}. \cite{allison2009a} also studied mass segregation in the ONC, finding a similar pattern, but with a threshold of $\approx 5$ $\msun$ below which there was no significant segregation instead. Monoceros R2 \citep{carpenter1997} and NGC 1983 \citep{sharma2007} show similar behavior. In this section, we investigate whether our simulations display this pattern of mass segregation as well.  
 
Following \cite{bressert2010}, we define the stellar density around a sink particle out to the $N$th neighbor as:
\beq
n_*(N) = \frac{N-1}{(4/3) \pi r_N^3},
\eeq where $r_N$ is the distance to the $N$th closest sink. The choice of $N$ is somewhat arbitrary; in what follows, we take $N = 9$ for all numerical results, and verify that our qualitative conclusions are not sensitive to this choice over the range $N = 4$ to $N = 20$. Likewise, we define the stellar surface density as 
\beq
N_*(N) = \frac{N - 1}{\pi r_N^2},
\eeq as this quantity is closer to what observers measure. For every star in our simulations, we compute $n_*$ and $N_*$ and plot these quantities versus the star mass $m_*$ in Figures \ref{fig:ms_volume} and \ref{fig:ms_surface}. There are two interesting features revealed in these plots. First, although the Strong and Weak runs have advanced to approximately the same time and have approximately the same number of stars, the stars in the Strong run tend to be found at higher stellar densities. The Strong run has 11 stars with $n_* = 1 \times 10^5$ stars pc$^{-3}$ or greater, while neither of the other runs do. The mean and median $n_*$ are higher in the Strong run as well (see Table \ref{table:mass_segregation}). This trend is also visible in Figures \ref{fig:Weakpanels} and \ref{fig:Strongpanels}, where the star formation appears more clustered in Strong than in Weak, in that roughly the same number of stars are confined to a smaller surface area. 
 
The second is that, with a few exceptions, all the stars with $m_* > 1$ $\msun$ are found in regions of relatively high stellar density. To make this more quantitative, we compute the mean and median values of $n_*$ twice, once for stars with $m_* < 1$ $\msun$, and again for stars with $m_* > 1 $ $\msun$. We do the same for the stellar mass density, $\rho_*$, defined as the total stellar mass within a distance $r_N$ around each star. The results are summarized in Table \ref{table:mass_segregation} for the Hydro, Weak, and Strong runs. We also show the combined properties of all three runs. With the exception of the Hydro run, which only has 2 stars with  $m_* > 1$ $\msun$, the mean $n_*$ for super-solar stars is larger than those of sub-solar stars by about a factor of $\approx 6$, while the median is larger by a factor of $\approx 30$. If we compare the stellar mass densities instead, the effect is even more pronounced. We have also indicated in Figures \ref{fig:ms_volume} and \ref{fig:ms_surface} the median value of $n_*$ for sub-solar stars in each run. In the Strong run, only one $> 1$ $\msun$ star lies in a region where the stellar surface density is below the median for all the $< 1$ $\msun$ stars in the run.  In the Weak run, none do. 

\begin{table*}
\begin{minipage}{175mm}
\caption{ \label{table:mass_segregation} Stellar surface density around $> 1$ $\msun$ versus $< 1$ $\msun$ stars \label{segregation}}
\centering
\begin{tabular}{@{}cccccccccccccc}
\hline
Name & $n_{*_{{50, <}}}$ & $n_{*_{{50, >}}}$ & $\bar n_{*_<}$ & $\bar n_{*_>}$ & $\rho_{*_{50,<}} $ & $\rho_{*_{50, >}} $ & $\bar \rho_{*,<}  $ & $\bar \rho_{*,>} $ & $p$-value \\
\hline
Hydro  &$2.0$& -- & 8.6 & -- & 0.5 & -- & 2.5 & -- & --   \\
Weak &$2.1$ & 74.1 & 9.0 & 64.6 & 0.5 & 117.1 & 8.5 & 96.7 & 5.3 $\times 10^{-5}$   \\
Strong &$8.2$ & 226.0 & 31.4 & 160.5 & 3.1 & 250.9 & 25.6 & 171.8 & 2.3 $\times 10^{-3}$    \\
All  &$2.7$& 74.1 & 16.8 & 91.6 & 0.7 & 113.7 & 12.5 & 110.0 & 2.6 $\times 10^{-6}$   \\
  \hline
 \end{tabular}

\medskip
Col.\ 1 : median number density for stars with $m_* < 1$ $\msun$ in units of 1000 stars per pc$^{3}$. Col.\ 2: same, but for stars with $m_* > 1$ $\msun$. Col. \ 3 and \ 4:  same, but the mean instead of the median. Col.\ 5 through 8: same as Col. \ 1 through 4, but for the stellar mass density in units of 1000 $\msun$ per pc$^{-3}$ instead of the number density. Col.\ 9: $p$-value returned by a K-S test comparing the $m_* < 1$ $\msun$ and $m_* > 1$ $\msun$ distributions.
\end{minipage}
\end{table*}

We also show in Table \ref{table:mass_segregation} the $p$-value associated with a two-sided Kolmogorov-Smirnov (K-S) test comparing the distributions of $n_*$ for $< 1$ $\msun$ and $> 1$ $\msun$ stars. For the Weak and Strong MHD cases, we can reject the null hypothesis that the two populations are drawn from the same underlying distribution at the 0.05\% and 5\% confidence levels, respectively. For the Hydro case, this number is not particularly meaningful, since there are only a couple of stars larger than 1 $\msun$. Finally, for the combined sample of all the stars formed in all three runs, the $p$-value that $< 1$ $\msun$ stars and $> 1$ $\msun$ stars have the same distribution is only $2.6 \times 10^{-6}$. 

Note that the same is not true if we repeat this procedure with a different mass threshold. For example, if we compare the distribution of $n_*$ around stars with $ 0.05$  $\msun < m_* < 0.4$ $\msun$ to that of stars with $ 0.4$ $\msun < m_* < 1.0$ $\msun$ (the threshold of $0.4 \msun$ was picked because it divides the stars in the mass range 0.05 to 1.0 $\msun$ into two groups of approximately equal mass), we get K-S $p$-values of 0.30, 0.59, and 0.56, for the Hydro, Weak, and Strong runs, respectively. In other words, the data for stars with $ 0.05$  $\msun < m_* < 0.4$ $\msun$ are consistent with being drawn from the same underlying distribution as those with $ 0.4$ $\msun < m_* < 1.0$ $\msun$. There appears to be a real difference in our simulation between $< 1$ $\msun$ stars, which are found at both low and high stellar density independent of mass, and $> 1$ $\msun$ stars, which are much more likely to be found at high $n_*$. 

Our threshold value of 1 $\msun$ is lower than the threshold for the ONC by about a factor of 3. It is perhaps not surprising that we do not agree with the ONC value quantitively, since the most massive star in our simulations is $\approx 5.2$ $\msun$, while $\theta^1$ Orionis C, the most massive member of the Trapezium, is $\approx 37$ $\msun$ \citep{kraus2009}. Nonetheless, we do reproduce the fact that beyond some threshold mass, stars are much more likely to form in the center of clusters. Interestingly, this same basic pattern has been observed for protostellar \emph{cores} as well. In a study of dense cores in the $\rho$ Ophiuchi cloud complex, \cite{stanke2006} found no mass segregation for starless cores with masses $\lesssim 1$ $\msun$, but the most massive cores were only found in the dense, inner region. Finally, although N-body processing can produce the mass segregation observed in the ONC on timescales comparable to the cluster age of a few Myr \citep[e.g.][]{allison2009b}, insufficient time (only 56,000 kyr) has elapsed for this effect to be important here. The mass segregation in our simulations is primordial, rather than dynamical. 
 
\subsection{Multiplicity}
\label{sec:multiplicity}

Finally, we consider the multiplicity of the stars formed in our simulations. To divide our star particles into gravitationally bound systems,  we use the algorithm of \cite{bate2009} \citep[see also][]{bate2012, krumholz2012}. We start with a list of all the stars in each simulation. For each pair, we compute the total center-of-mass frame orbital energy. We then replace the most bound pair with a single object that has the total mass, net momentum, and center-of-mass position of its constituent stars. We repeat this procedure until there are no more bound pairs. The only exception is that we do not create systems with more than four stars - if combining the most bound pair of objects would create a system with 5 or more stars, we combine the next most bound pair instead \footnote{Our qualitative results are essentially the same if we choose a slightly different limit.}. At the end of this process, there are no more pairs that can be combined, either because they are not gravitationally bound or because combining them would result in more than 4 stars in a system. 

We then compute both the multiplicity fraction MF \citep{hubber2005, bate2009, krumholz2012}:
\beq
\label{multiplefrac}
\rm{MF} = \frac{B + T + Q}{S + B + T + Q},
\eeq where $S$, $B$, $T$, and $Q$ are the numbers of single, binary, triple, and quadruple star systems, and the companion fraction \citep[e.g.][]{haisch2004}:
\beq
\label{multiplefrac}
\rm{CF} = \frac{B + 2 T + 3 Q}{S + B + T + Q},
\eeq which is the number of companions per system. The CF is often reported in observations, but the MF has the desirable property that it does not change if a high-order system is re-classified as a binary or vice-versa.

\begin{table}
\caption{ \label{table:multiplicity} Multiple star systems}
\centering
\begin{tabular}{@{}ccccccccc}
\hline
Name & $S$ & $B$ & $T$ & $Q$ & $MF$ & $CF$ & \\
\hline
Hydro  &139 & 10 & 2 & 3 & 0.10 & 0.15   \\
Weak &65 & 4 & 5 & 5 & 0.18 & 0.37   \\
Strong &66 & 9 & 2 & 8 & 0.22 & 0.44   \\
Hydro$_r$ & 138 & 12 & 3 & 1 & 0.10 & 0.14 \\
Weak$_r$ & 67 & 8 & 3 & 1 & 0.15 & 0.22 \\
Strong$_r$ & 67 & 11 & 5 & 2 & 0.21 & 0.32 \\
  \hline
 \end{tabular}

\medskip
Rows 1-3 - all companions, regardless of separation. Rows 4-6 - not counting companions with separations $< 200$ or $> 4500$ AU.
\end{table}

The results of this calculation for our three runs are shown in Table \ref{table:multiplicity}. We find that there is a clear trend towards more multiplicity with stronger magnetic fields. This likely related to the phenomenon discussed in \ref{sec:globalevolution} and \ref{sec:mass_segregation}. Stellar clustering is more dense in the Strong run than the others, with roughly the same number of stars packed into a smaller volume, so the availability of potential partners tends to be greater in run Strong.

At this point, we mention a few caveats of this analysis. First, in our simulations, we are only marginally able to resolve binaries closer than our sink accretion radius of $r_{\rm{sink}} = 4 \Delta x_f \approx 184$ AU. Due to the way our sink particle algorithm works, binaries are unable to form within this distance. Likewise, binaries where one of the partners forms outside of this distance but falls in before exceeding the minimum merger mass of $0.05 \msun$ would be counted as a single star in these simulations. Stars that form outside $r_{\rm{sink}}$ \emph{and} exceed this threshold before falling in are able to form binary systems closer than $r_{\rm{sink}}$. However, there is an additional problem, which is that sink-sink gravitational forces are softened on a scale of $0.25$ $ \Delta x_f \approx 11.5$ AU and gas-sink gravitational forces on a scale of $\Delta x_f \approx$ 46 AU. We thus have only a limited ability to resolve binaries with separations $\lesssim 200$ AU. 

Most main-sequence, solar-type stars are members of binaries \citep{duquennoy1991}, and young stellar objects such as T Tauri stars \citep{duquennoy1991, patience2002} and Class I sources \citep{haisch2004, duchene2007, connelley2008, duchene2013} have an even greater tendency to be found in bound multiple systems. The sink particles in our simulations are all still embedded, but have begun to heat up their immediate surroundings to temperatures high enough to radiate in the infrared, and thus are most analogous to Class I sources. Observations of multiplicity among Class I objects have difficulty detecting both very tight and very wide binaries, and thus generally report a \emph{restricted} companion fraction - that is, the CF counting only companions within some range of projected separations. For instance, \cite{connelley2008} found a CF of 0.43 for Class I sources in nearby star-forming regions within the range of 100 to 4500 AU, while the high-resolution observations of \cite{duchene2007} found CF $= 0.47$ for 14 to 1400 AU. To compare against these observations, we therefore compute the restricted companion fractions in our simulations in the range 200 - 4500 AU. The lower limit of 200 AU restricts the analysis to binaries that are resolved at our grid resolution. The results of this calculation are also shown in Table \ref{table:multiplicity}. 

The main effect of restricting our analysis to companions in the range of 200 - 4500 AU is to re-classify triple and quadruple star systems as binaries. This is because the higher-order multiples in our simulations are generally hierarchical, with, say, a wide companion orbiting a tighter binary system. Looking for companions only between 200 - 4500 AU misses many of these partners. This effect makes no difference for the MF, but can change the CF significantly, particularly in the Weak and Strong runs, which without restriction had many triple and quadruple systems. Discounting the companions in the range 100 - 200 AU, \cite{connelley2008} found a CF of 0.33 in the range 200 - 4500 AU. This is quite close to our Strong run, in which the CF restricted to the same range is 0.32.   

Additionally, like \cite{bate2012} and \cite{krumholz2012}, the multiplicity fraction in our calculations is a strong function of primary mass, $m_p$. If we consider only systems in which the primary star exceeds $1$ $M_{\odot}$, we find that there are 3 such systems in run Hydro, 4 in Weak, and 5 in Strong. Only one of these systems, however, is single. This suggests that the trend for higher multiplicity at higher primary masses, which is well-observed for main-sequence stars, may already be in place during the Class I phase. This is likely related to the phenomenon discussed in section \ref{sec:mass_segregation}: that stars more massive than $\approx 1$ $\msun$ are much more likely than average to form in regions of high stellar density. Thus, more massive stars tend to form in regions where there are more potential partners for forming binary and other higher multiple systems. Another potential mechanism behind the strong mass dependence of the multiplicity fraction - that more massive stars have higher accretion rates and thus are more likely to be subject to disk fragmentation \citep{kratter2010} - is unlikely to be responsible for the trend in our simulations, simply because at a resolution $\Delta x_f \approx 46$ AU, we are not able to resolve any disk physics. 

 \section{Discussion}
 \label{sec:discussion}
 
We find that the magnetic field influences most aspects of cluster formation and early evolution, including the star formation rate, the degree of fragmentation, the median fragment mass, the multiplicity fraction, and the typical stellar density in the cluster. However, the magnitude of these effects are rather modest at $\mu_{\Phi} = 2$, differing from the pure Hydro case at roughly the factor of $2$ level. While in general our magnetized runs, particularly the $\mu_{\Phi} = 2$ run, compare favorably with observations compared to our non-magnetized run, the differences are not dramatic. 

In \ref{sec:IMF}, we compared the properties of our simulated protostars to the \cite{chabrier2005} stellar IMF, finding that while our simulations agreed well with the log-normal functional form, the characteristic masses were lower than the \cite{chabrier2005} value of $m_c = 0.2$ $M_{\odot}$ by a factor of 2-4, depending on the magnetic strength. This is to be expected, since even at late times our population of sink particles includes newly-formed objects that are not close to their final masses, and most of the more evolved objects are still accreting significantly. \cite{krumholz2012}, which used initial conditions similar to our Hydro run but with mixed solenoidal/compressive forcing, found good agreement with the \cite{chabrier2005} system IMF ($m_c = 0.25$) and the \cite{dario2012} IMF for the Orion Nebula Cluster ($m_c = 0.35$ $\msun$). The typical protostar in \cite{krumholz2012} was thus significantly larger than the typical protostar in this work. 

This difference is likely to due to the varying degree of effectiveness of radiative feedback in our two simulations. Because the star formation rate was higher in \cite{krumholz2012}, there was considerably more protostellar heating, which pushed the characteristic fragmentation scale to higher masses. If we compare our Figures \ref{fig:PhaseMass} and \ref{fig:PhaseTime} against Figure 10 of \cite{krumholz2012}, we see that there is significantly less gas that has been heated above the background temperature in our Hydro run than in run TuW of \cite{krumholz2012}. Quantitatively, \cite{krumholz2012} report that 7\% of the gas in run TuW is hotter than 50 K at the latest time available. The corresponding values for the most evolved stage of our simulations shown in Figure \ref{fig:PhaseMass} are 0.3\% 0.3\%, and 0.1\% for Hydro, Weak, and Strong, respectively. This difference in protostellar heating made the particle masses in \cite{krumholz2012} agree well with the IMF, even though the majority of the sinks were still accreting. With the lower star formation rates in this work, our median sink particle mass drops to something more characteristic of a protostellar mass function, rather than an IMF (see Sec. \ref{sec:PMF}). 

However, our simulations confirm the result of \cite{krumholz2012} that when turbulent initial conditions are treated self-consistently, the population of sink particles can approach a steady-state mass distribution (Figure \ref{fig:ks}). The ``overheating problem" identified by \cite{krumholz2011} for simulations in which  star formation is too rapid does not occur here, and the characteristic stellar mass is relatively stable with time. 

The first simulations of star cluster formation in turbulent molecular clouds to include both magnetic fields and radiative feedback are the smoothed-particle hydrodynamics (SPH) simulations of \cite{price2009}. \cite{price2009} found that the median protostellar mass tended to \emph{decrease} with increasing magnetic field strength in their radiative calculations, the opposite of the trend reported here, although they cautioned that larger simulations that form more sink particles were necessary before drawing firm conclusions. One potential reason for the difference between our result and \cite{price2009} is that the star formation in \cite{price2009} occurs in the context of a globally collapsing structure, in which stars form in the center and accrete in-falling gas before getting ejected by N-body interactions. Because this rate of infall is lower with stronger $B$ fields, the typical particle accreted less material before being ejected. In contrast, in our simulations there is no global infall. The typical star forms from a core that results from turbulent filament fragmentation, and the magnetic field increases the typical fragment mass.    
 
 \section{Conclusions}
 \label{sec:conclusions}
 
We have presented a set of simulations of star-forming clouds designed to investigate the effects of varying the magnetic field strength on the formation of star clusters. We find that the magnetic field strength influences cluster formation in several ways. First, the magnetic field lowers the star formation rate by a factor of $\approx 2.4$ over the range $\mu_{\Phi} = \infty$ to $\mu_{\Phi} = 2$, in good agreement with previous studies \citep{price2009, padoan2011, federrath2012}. Second, it also suppresses fragmentation, reducing the number of sink particles formed in our simulations by about a factor of $\approx 2$ over the same range of $\mu_{\Phi}$. This too is in good agreement with previous work \citep{hennebelle2011, federrath2012}. 

The magnetic field also tends to increase the median sink particle mass, again by a factor of about 2.4 over the range of $\mu_{\Phi} = \infty$ to $\mu_{\Phi} = 2$. Even at $\mu_{\Phi} = 2$, however, the median sink mass is still lower than the value for the \cite{chabrier2005} IMF by about 40\%, likely because our sinks are still accreting at the time we stop our calculations. On the other hand, our $\mu_{\Phi} = 2$ calculation is statistically consistent with both the two-component turbulent core and the two-component competitive accretion protostellar mass functions from \cite{mckeeoffner2010} and \cite{mckeeoffner2011}. In contrast, the pure Hydro simulation does not agree well with either the \cite{chabrier2005} IMF or any of the PMFs in \cite{mckeeoffner2010}.  

We also find that the accretion rates onto the most massive stars in our simulations (about $\sim 2 - 5$ $\msun$) are well-described by the TC model. We have confirmed that these accretion rates depend only weakly on the magnetic field strength, as predicted by \cite{mckee2003}.

We examined the magnetic field geometry in our simulations at the $\sim 0.005$ pc scale. In the Strong field case, the field geometry agrees well with observations of low-mass \citep{girart2006, rao2009} and high-mass \citep{girart2009, tang2009} star-forming cores, but the magnetic field lines are often quite disordered in the Weak run. If, as suggested by \cite{crutcher2010}, $\approx 10$\% of cores have $\mu_{\Phi} \gtrsim 10$, then we would expect a similar fraction of observed cores to reveal disordered fields at the $\sim 3000$ AU scale.

Many of the stars in our simulations are members of bound multiple systems, and our Strong field run in particular agrees well with observations of multiplicity in Class I sources over the range of 200 - 4500 AU \citep{connelley2008}. We find a trend towards increased multiplicity with magnetic field strength that is likely explained by the fact that star formation is more clustered in the Strong run than others, since at $\mu_{\Phi} = 2$ much of the volume is prevented from collapsing gravitationally. Our simulations also reproduce the fact, observed in main-sequence stars, that more massive stars are more likely to be found in multiple systems than their lower-mass counterparts.

Finally, all our simulations exhibit a form of primordial mass segregation like that observed in the ONC, in which only the most massive stars are more likely than average to be found in regions of high stellar density. 
  
\section*{Acknowledgments} 
A.T.M. wishes to acknowledge the anonymous referee for a thorough report that improved the paper, and to thank Andrew Cunningham and Eve Lee for fruitful discussions while the paper was in preparation. Support for this work was provided by NASA through ATP grant NNX13AB84G (R.I.K., M.R.K. and C.F.M.) and a Chandra Space Telescope grant (M.R.K.); the NSF through grants AST-0908553 and NSF12-11729 (A.T.M., R.I.K. and C.F.M.) and grant CAREER-0955300 (M.R.K.); an Alfred P. Sloan Fellowship (M.R.K); and the US Department of Energy at the Lawrence Livermore National Laboratory under contract DE-AC52-07NA27344 (A.J.C. and R.I.K.) and grant LLNL-B602360 (A.T.M.). Supercomputing support was provided by NASA through a grant from the ATFP. We have used the YT toolkit \citep{yt} for data analysis and plotting. 

\bibliographystyle{mn2e}
\bibliography{mn-jour,bibliography}

\end{document}